\documentclass[aps,pra,amsmath,reprint,floatfix,onecolumn,notitlepage,superscriptaddress]{revtex4-1}

\usepackage{microtype} 
\usepackage{amssymb}
\usepackage{graphicx}
\usepackage{enumerate}
\usepackage{bm}

\RequirePackage[
  hyperindex,colorlinks,bookmarksnumbered,
  plainpages=true,pdfstartview=FitH]{hyperref}
\hypersetup{linkcolor=blue,urlcolor=blue,citecolor=blue} 
\usepackage{hyperref}
\usepackage[all]{hypcap}

\newcommand{\ED}{.}
\newcommand{\EC}{,}

\newcommand{\p}{^{\prime}}
\newcommand{\pp}{^{\phantom\prime}}
\newcommand{\ppp}{^{\protect\phantom\prime}} 

\renewcommand{\vec}[1]{\bm{#1}}

\begin{document}

\title{Derivation of exact flow equations from the self-consistent parquet relations}
\author{Fabian B.~Kugler}
\affiliation{Max-Planck-Institut f\"ur Quantenoptik, Hans-Kopfermann-Str.~1, 85748 Garching, Germany}
\affiliation{Physics Department, Arnold Sommerfeld Center for Theoretical Physics, and Center for NanoScience, Ludwig-Maximilians-Universit\"at M\"unchen, Theresienstr.~37, 80333 Munich, Germany}
\author{Jan von Delft}
\affiliation{Physics Department, Arnold Sommerfeld Center for Theoretical Physics, and Center for NanoScience, Ludwig-Maximilians-Universit\"at M\"unchen, Theresienstr.~37, 80333 Munich, Germany}

\date{\today}

\begin{abstract}
We exploit the parquet formalism to derive exact flow equations for
the two-particle-reducible four-point vertices, the self-energy, and typical response functions,
circumventing the reliance on higher-point vertices.
This includes a concise, algebraic derivation of the multiloop flow equations, which have previously been obtained by diagrammatic considerations. 
Integrating the multiloop flow for a given input of the totally irreducible vertex is equivalent to solving the parquet equations with that input. 
Hence, one can tune systems from solvable limits to complicated situations by variation of one-particle parameters, staying at the fully self-consistent solution of the parquet equations throughout the flow.
Furthermore, we use the resulting differential form of the Schwinger-Dyson equation for the self-energy to demonstrate one-particle conservation of the parquet approximation and to construct a conserving two-particle vertex via functional differentiation of the parquet self-energy. Our analysis gives a unified picture of the various many-body relations and exact renormalization group equations.
\end{abstract}

\maketitle

\section{Introduction}

The many-body problem of nonrelativistic quantum field theory is equipped with a well-known set of exact equations for its correlation functions \cite{Hedin1965,Bickers2004}. If these self-consistent many-body relations are expressed in their energy-momentum representation, they interrelate the different correlation functions between all energy scales, often involving integrations over all energy-momenta.
However, a typical feature of interacting quantum many-body systems is that their relevant energy scales span several orders of magnitude. Conventional perturbative approaches or approaches that directly work with the self-consistent many-body relations treat all energy scales at once---they are therefore prone to inaccuracies and often plagued by infrared divergences.
A very successful approach to such systems is instead given by the renormalization group (RG) technique which treats energy scales successively, starting from high ones and progressing towards lower ones \cite{Wilson1975}.

The simplest realization of such a RG scheme considers the renormalization of effective couplings in analogy to Anderson's poor man's scaling \cite{Anderson1970}. There, the successive treatment of high-energy degrees of freedom is encoded in the evolution of running coupling constants.
Since then, quantum-field-theoretical RG techniques have seen great development. A widely used, modern formulation is given by the functional RG (fRG), which allows one to study the flow of all coupling ``constants'' in their full functional dependence \cite{Metzner2012,Kopietz2010}. The respective couplings are nothing but the (field-theoretical) vertex functions; hence, the fRG can be directly applied to microscopic models. 

The fRG flow is based on an exact functional flow equation for the generating functional of the (one-particle-irreducible) vertex functions \cite{Wetterich1993}. If this flow equation is expanded in terms of the vertices, one obtains an infinite hierarchy of flow equations, where, in order to compute the flow of an $n$-point vertex, knowledge about the other vertices up to the $n+2$-point vertex is required. The obvious way of truncating the hierarchy by disregarding higher-point vertices has led to a variety of successful applications of the fRG. However, one often wants to extend the usage of fRG beyond the validity of this approximation, and, in cutting-edge algorithmic development, this form of truncation may indeed be an exceedingly severe approximation. 

In fact, considering a system of, say, interacting electrons, possibly subject to external fields,
one may ask why it is necessary to include six- and higher-point vertices, i.e., \textit{effective} interactions between three and more particles, if one is ultimately interested in one- and two-particle properties of the system.
Although the fRG hierarchy of flow equations and also the hierarchy of Schwinger-Dyson equations (SDEs, or equations of motion) \cite{Veschgini2013} interrelate all $n$-point vertices, the fundamental interaction is only of the (one- and) two-particle type; thus, it should suffice to work on the one- and two-particle level.
Fortunately, a many-body framework that provides a complete description on the one- and two-particle level is available; it is the parquet formalism \cite{Bickers2004,Roulet1969}.

The main idea of the approach presented in this paper is to apply the RG point of view neither to the generating functional of vertices \cite{Wetterich1993} nor to the hierarchy of SDEs \cite{Veschgini2013} but to the self-consistent many-body relations of the parquet formalism. Exploiting the organizational structure of the parquet formalism allows us to circumvent the inclusion of higher-point vertices and to freely navigate between different two-particle channels.
Inspired by the fRG framework, we induce an internal scale dependence by using a scale-dependent propagator $G^{\Lambda}$ that suppresses low-energy degrees of freedom and recovers the original theory at a final value $\Lambda_f$.
It should be noted that this differs in technical aspects from more traditional RG schemes \cite{Wilson1975,Shankar1994}, which, instead of solely using a scale-dependent propagator, restrict all involved energy-momenta to decreasing energy-momentum shells (often referred to as ``mode elimination'').
Here, we simply substitute $G \to G^{\Lambda}$ in the well known many-body relations and study the behavior of the solution to these equations upon varying $\Lambda$.

As a result, we derive exact flow equations for
the two-particle-reducible four-point vertices, the self-energy, and response functions.
This provides a concise, algebraic derivation of the multiloop fRG (mfRG) flow equations, which have previously been obtained using diagrammatic arguments \cite{Kugler2018,Kugler2018a,Tagliavini2018}.
Our analysis also reveals how one can perform such multiloop flows beyond the parquet approximation (PA), thus including higher-order expressions for the totally irreducible vertex. Moreover, we establish an intimate connection between the functional derivative of the self-energy and the fRG flow equation for the self-energy: the latter constitutes an integration of the former along a specific path in the space of theories.

On a slightly different note, we use our approach to address fundamental questions of (traditional) parquet theory (i.e., without an explicit RG treatment): 
On the one hand, we demonstrate that the parquet self-energy can be obtained from the Schwinger-Dyson equation (SDE) using either of two possible orderings of the bare and full vertex. According to Baym and Kadanoff \cite{Baym1961}, it then follows that the PA fulfills one-particle conservation laws.
On the other hand, we give an explicit construction to obtain a new, conserving vertex from the parquet self-energy, equivalent to taking the functional derivative. 
This construction not only allows one to quantify the degree to which the PA violates two-particle conservation laws.
It can also be used to modify the PA, which fulfills the SDE but violates two-particle conservation, to obtain a fully conserving solution, albeit violating the SDE. As we show in the appendix, a fulfillment of both the SDE and the functional-derivative relation necessarily amounts to the exact solution of the many-body problem, in agreement with a result by Smith \cite{Smith1992}.

The paper is structured as follows. In Section \ref{sec:vertex}, we first focus (as is typical for RG approaches) on the effective interactions: we derive flow equations for the two-particle-reducible four-point vertices based on the parquet formalism, assuming the one-particle propagator to be given. 
Then, in Section \ref{sec:self-energy}, we complement the flow of the four-point vertex by the flow of the self-energy, considering the various relations at hand. 
In Section \ref{sec:conservation}, we use our approach to discuss conservation properties of the PA.
Finally, in Section \ref{sec:response}, we derive (dependent) flow equations for response functions, i.e., three-point vertices and suceptibilities, used to study collective excitations.
In Section \ref{sec:conclusion}, we summarize our results.
\section{Derivation of the vertex flow}
\label{sec:vertex}
\subsection{Preliminaries}
We consider a general theory of interacting fermions,
defined by the action
\begin{align}
S & = - \sum_{x\p, x} \bar{c}_{x\p} \big[ (G_0)^{-1} \big]_{x\p, x} c_{x} 
- \tfrac{1}{4} 
\!\!\!\! \sum_{x\p,x,y\p,y} \!\!\!\!
\Gamma_{0;x\p,y\p;x,y} \bar{c}_{x\p} \bar{c}_{y\p} c_{y} c_{x}
\EC
\label{eq:action}
\end{align}
with a bare propagator $G_0$ and 
a bare four-point vertex $\Gamma_0$, which is
antisymmetric in its first and last two arguments.
The index $x$ denotes all quantum numbers of the Grassmann field $c_x$.
Correlation functions of fields, corresponding to time-ordered
expectation values of operators, are given by the functional integral
\begin{equation}
\langle c_{x_1} \cdots \bar{c}_{x_n} \rangle = 
\frac{1}{Z}
\int \! \mathcal{D}[\bar{c}] \mathcal{D}[c] \,
c_{x_1} \cdots \bar{c}_{x_n} e^{-S}
\EC
\end{equation}
where $Z$ ensures normalization, such that $\langle 1 \rangle=1$.
Two-point correlation functions are represented by the full propagator 
$G_{x, x\p} = - \langle c_{x} \bar{c}_{x\p} \rangle$;
four-point correlation functions can be expressed via the full
(one-particle-irreducible) 
four-point vertex $\Gamma$:
\begin{align}
\langle c_{x_1\pp} c_{x_2\pp} \bar{c}_{x_2\p} \bar{c}_{x_1\p} \rangle
& =
G_{x_1\pp x_1\p} G_{x_2\pp x_2\p} - G_{x_1\pp x_2\p} G_{x_2\pp x_1\p}
+
G_{x_1\pp y_1\p} G_{x_2\pp y_2\p}
\Gamma_{y_1\p, y_2\p; y_1\pp, y_2\pp}
G_{y_1\pp x_1\p} G_{y_2\pp x_2\p}
\ED
\label{eq:four-point_correlator}
\end{align}
The notation given so far is identical to the one in Ref.\ \onlinecite{Kugler2018a};
all formulae further needed in this paper are defined in Appendix \ref{sec:matrixnotation}.
In the following derivation of flow equations, we use a compact notation of contractions and need not write quantum numbers (such as $x, x'$, etc.) explicitly.
\subsection{Parquet equations for the four-point vertex}
\begin{figure}[t]
\includegraphics[width=.85\textwidth]{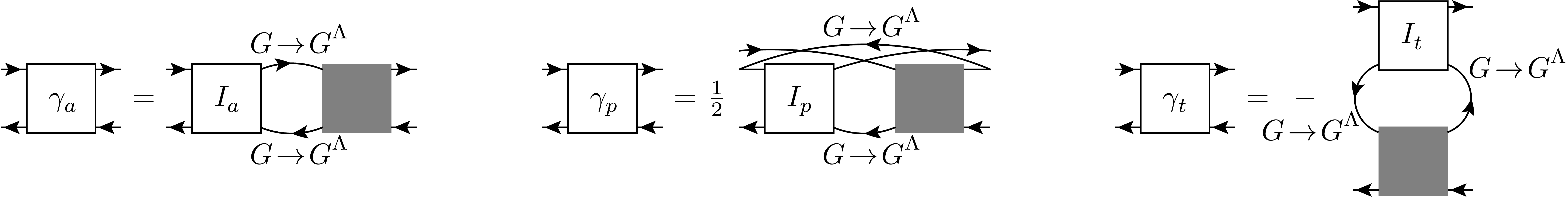}
\caption{%
The Bethe-Salpeter equations for the channels $r=a,p,t$ are solved in an RG approach by introducing a scale ($\Lambda$) dependence to the propagators connecting the vertices. Consequently, 
$\gamma_r$, $\Gamma$, and $I_r$
inherit a scale dependence while the totally irreducible vertex, $R$, remains as given input.
(See Appendix \ref{sec:matrixnotation} for details on the diagrammatic notation.)
As prime example for the scale dependence, one can multiply the frequency-dependent propagator by a step function, $G^{\Lambda}(\omega) = \Theta(|\omega|-\Lambda) G(\omega)$, such that the many-body relations are trivially solved at $\Lambda_i=\infty$ 
and reproduce the desired solution at $\Lambda_f=0$.%
}
\label{fig:bse}
\end{figure}

The fRG flow equation for the four-point vertex, $\Gamma^{(4)} \equiv \Gamma$, 
contains the six-point vertex, $\Gamma^{(6)}$, which
poses great difficulty for a numerical treatment.
Similarly, the SDE (equation of motion) for $\Gamma$ contains $\Gamma^{(6)}$ and therefore is likewise impractical.
To circumvent the calculation of $\Gamma^{(6)}$,
we revert to the parquet formalism \cite{Bickers2004,Roulet1969}, which provides 
self-consistent equations for the two-particle-reducible contributions to the four-point vertex $\Gamma$
but assumes as input a given, totally irreducible four-point vertex $R$.
In a diagrammatic expansion, $R$ is given by the bare vertex, $\Gamma_0$, with corrections starting at fourth order. The famous \textit{parquet approximation} \cite{Yang2009,Tam2013,Li2016} (see Section \ref{sec:conservation}) consists of using $R=\Gamma_0$ and allows one to sum up all leading logarithmic contributions in logarithmically divergent perturbation theories \cite{Roulet1969,Abrikosov1965}. Importantly, however, the parquet equations can be used more generally as an exact classification of all diagrams of the four-point vertex.
In the parquet formalism, one decomposes the full four-point vertex, $\Gamma$, into the totally irreducible vertex, $R$, and the three two-particle-reducible vertices $\gamma_r$, $r \in \{a,p,t\}$
\footnote{
Our nomenclature follows the seminal application of the parquet equations
to the X-ray-edge singularity by Roulet et al.\ \cite{Roulet1969}.
While we use $\Gamma$, $R$, $\gamma_r$, and $I_r$ for the full, totally irreducible, two-particle-reducible, and -irreducible vertices, respectively, another common choice \cite{Rohringer2017,Rohringer2012,Wentzell2016} is given by $F$, $\Lambda$, $\Phi_r$, $\Gamma_r$, respectively.
Similarly, a common notation \cite{Rohringer2017,Rohringer2012,Wentzell2016}
for the channels $a, p, t$
is $ph, pp, \overline{ph}$, referring to
the (longitudinal) particle-hole, the particle-particle,
and the transverse (or vertical) particle-hole channel, respectively.
One also finds the labels $x, p, d$ in the literature \cite{Jakobs2010}, 
referring to the
so-called exchange, pairing, and direct channel, respectively.}%
\nocite{Roulet1969,Rohringer2017,Rohringer2012,Wentzell2016,Jakobs2010}%
. Diagrams belonging to $\gamma_r$ are reducible in channel $r$, i.e., they can be separated into two parts by cutting two $a$ntiparallel, $p$arallel, or $t$ransverse antiparallel lines, respectively. Diagrams that cannot be separated in this way belong to $R$. (For exemplary diagrams, see Fig.~\ref{fig:indices} in Appendix \ref{sec:matrixnotation}.) While the $\gamma_r$ are subject to further equations, this set of coupled equations closes only for a fixed choice of $R$.
Let us assume a given expression for the totally irreducible vertex, $R$.
Furthermore, we will for now assume the one-particle propagator, $G$, to be given;
computation of $G$ via the self-energy will be discussed later.
The \textit{parquet equations}, involving the
two-particle-reducible vertices, $\gamma_r$, and two-particle-irreducible vertices, $I_r$, read
\begin{subequations}
\begin{align}
\Gamma 
= 
R + \textstyle\sum_r \gamma_r 
\EC 
&  
\quad 
I_r 
= 
\Gamma - \gamma_r
= R + \gamma_{\bar{r}}
\EC
\label{eq:parquet_decomposition}
\\
\gamma_r
&
= 
I_r \circ \Pi_r \circ \Gamma
\ED
\label{eq:bse}
\end{align}
\label{eq:parquet}%
\end{subequations}
For given $R$, these equations must be solved self-consistently to obtain the appropriate reducible vertices, $\gamma_r$, that complement the full vertex, $\Gamma$.
In Eq.~\eqref{eq:parquet_decomposition}, we use the notation $\bar{r}$ for the complementary channel of a given channel $r$, such that $\gamma_{\bar{r}} = \sum_{r'\neq r} \gamma_{r'}$.
The \textit{Bethe-Salpeter equation} (BSE) \eqref{eq:bse} describes two vertices, $I_r$ and $\Gamma$, connected by a bubble, $\Pi_r$, of two dressed propagators in channel $r$ (see also Fig.~\ref{fig:bse}).
This bubble of vertices can be expressed as a matrix multiplication (given a suitable parametrization depending on the channel $r$, cf.\ Appendix \ref{sec:matrixnotation}), as indicated by
the symbol $\circ$ attached to $\Pi_r$.
Note that $\Pi_p$ and $\Pi_t$ implicitly contain a factor of $1/2$ and $(-1)$, respectively.
In the following, we list relations that can be easily deduced from the parquet equations \eqref{eq:parquet} and will be used repeatedly in the derivations of flow equations.
The combination of Eqs.~\eqref{eq:parquet_decomposition} and \eqref{eq:bse} directly yields
$\Gamma = I_r + I_r \circ \Pi_r \circ \Gamma$ (for all channels $r$). 
Exploiting the multiplicative structure, we can isolate $\Gamma$ on the l.h.s.\ to obtain the \textit{inverted BSE},
\begin{equation}
\Gamma = I_r + I_r \circ \Pi_r \circ \Gamma
\quad
\Leftrightarrow
\quad
\Gamma = (1 - I_r \circ \Pi_r)^{-1} \circ I_r
\ED
\label{eq:ibse}
\end{equation}
A further straightforward manipulation yields an \textit{extended BSE},
\begin{equation}
1 + \Gamma \circ \Pi_r
=
1 +
(1 - I_r \circ \Pi_r)^{-1} \circ (I_r \circ \Pi_r - 1 + 1)
= 
(1 - I_r \circ \Pi_r)^{-1}
\ED
\label{eq:ebse}
\end{equation}
Using the inverted BSE \eqref{eq:ibse}, one directly sees (by isolating $\gamma_r$) that the order of the vertices in the BSE \eqref{eq:bse} is irrelevant:
\begin{equation}
\gamma_r 
= I_r \circ \Pi_r \circ \Gamma
= I_r \circ \Pi_r \circ (I_r + \gamma_r)
\quad
\Leftrightarrow
\quad
\gamma_r
= (1-I_r \circ \Pi_r)^{-1} \circ I_r \circ \Pi_r \circ I_r
= \Gamma \circ \Pi_r \circ I_r
\ED
\label{eq:rbse}
\end{equation}

\subsection{Flow of the four-point vertex}
The central aspect of our RG treatment is incorporated by attaching a scale ($\Lambda$) dependence 
to the propagator, $G \to G^{\Lambda}$, appearing in the self-consistent many-body relations. The physical picture is that $\Lambda$ separates high- and low-energy degrees of freedom, and by using $G^{\Lambda}$ we allow for successive renormalization of the low-energy ($<\Lambda$) theory by high-energy ($>\Lambda$) degrees of freedom as $\Lambda$ is decreased. However, one can also simply consider $\Lambda$ as some additional dependence in the propagators connecting the vertices in the BSEs:
$G \rightarrow G^{\Lambda}$, $\Pi_r \rightarrow \Pi_r^{\Lambda}$ (cf.\ Fig.~\ref{fig:bse}). 
Hence, the reducible vertices $\gamma_r^{\Lambda}$---and consequently $\Gamma^{\Lambda}$ and $I_r^{\Lambda}$---%
will inherit a scale ($\Lambda$) dependence, obtained from the requirement that the parquet relations be fulfilled for each value of $\Lambda$,
while $R$ remains as given input.
The scale dependence is auxiliary in the sense that we are ultimately interested in the fully renormalized theory: we are interested in $\gamma_r^{\Lambda_f}=\gamma_r$ where (at the final scale) $G^{\Lambda_f}=G$.
Suppose we know the vertices at the initial scale, i.e., we can solve the BSEs using $G^{\Lambda_i}$. Then, we can obtain $\gamma_r^{\Lambda_f}$ by solving a differential equation specified by the initial condition together with the flow $\partial_{\Lambda} \gamma_r^{\Lambda} \equiv \dot{\gamma}_r^{\Lambda}$, which is induced by the scale dependence of $G^{\Lambda}$ in the BSEs.
We remark that it is natural to exclude the totally irreducible vertex $R$ from the renormalization flow, as it constitutes precisely the part of the vertex that cannot be constructed iteratively and therefore does not have a flow equation that allows for an efficient (i.e., iterative one-loop) calculation.
\begin{figure}[t]
\includegraphics[width=.64\textwidth]{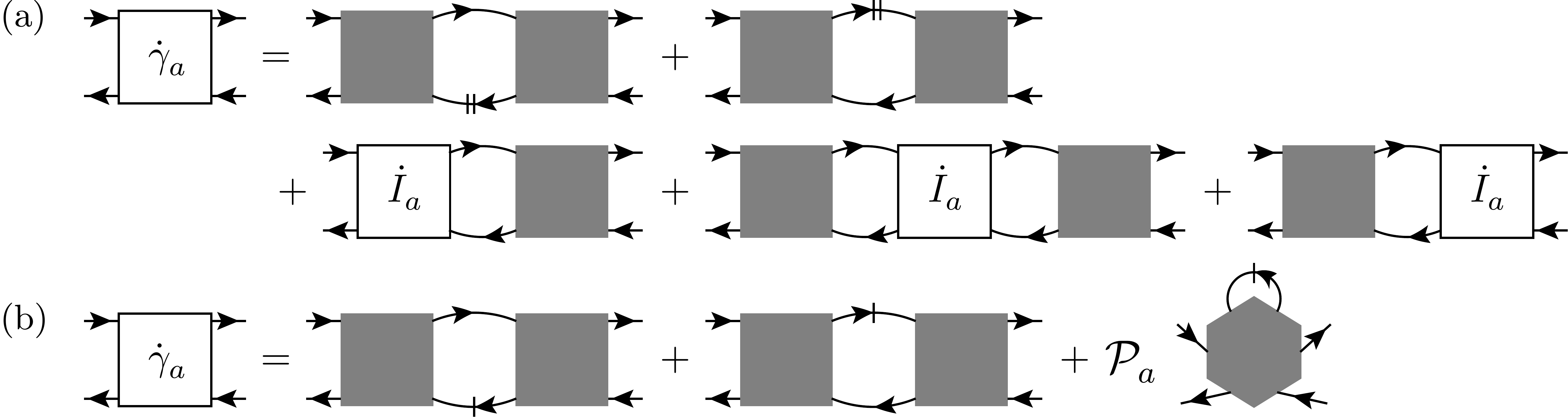}
\caption{%
(a)
Exact mfRG flow equation for the reducible vertex $\gamma_a$, involving the differentiated propagator, $\dot{G}$, (line with two vertical dashes) and the differentiated irreducible vertex given by $\dot{I}_r = \sum_{r'\neq r} \dot{\gamma}_{r'} \equiv \dot{\gamma}_{\bar{r}}$ (as $\dot{R}=0$ in our construction).
(b)
Exact fRG flow equation for $\gamma_a$ involving the single-scale propagator, $S=\partial_{\Lambda}G_{|_{\Sigma=\textrm{const}}}$, (line with one vertical dash) and the six-point vertex, whose contribution is (for conceptual purposes) reduced to the part reducible in the $a$ channel via the projector $\mathcal{P}_a$.%
}
\label{fig:gamma_flow}
\end{figure}

\subsubsection{Flow equation}
To find the scale dependence of the two-particle-reducible vertices, $\gamma_r$,
we start by differentiating the BSEs w.r.t.\ $\Lambda$
(suppressing the $\Lambda$ dependence to lighten the notation) according to the product rule and decomposing the full vertex via the parquet equation \eqref{eq:parquet_decomposition}:
\begin{align}
\dot{\gamma}_r
&
=
I_r \circ \dot{\Pi}_r \circ \Gamma 
+ \dot{I}_r \circ \Pi_r \circ \Gamma  
+ I_r \circ \Pi_r \circ 
\dot{\Gamma}
\nonumber \\
&
=
I_r \circ \dot{\Pi}_r \circ \Gamma 
+ \dot{I}_r \circ \Pi_r \circ \Gamma  
+ I_r \circ \Pi_r \circ 
(\dot{I}_r + \dot{\gamma}_r)
\ED
\label{eq:gamma_dot}
\end{align}
Similar to the manipulations in Eq.~\eqref{eq:rbse}, we bring $\dot{\gamma}_r$ to the l.h.s.\ and subsequently multiply by $(1-I_r \circ \Pi_r)^{-1}$ from the left. 
According to the inverted BSE \eqref{eq:ibse}, we get
\begin{align}
\dot{\gamma}_r
&
=
\Gamma \circ \dot{\Pi}_r \circ \Gamma
+
(1-I_r \circ \Pi_r)^{-1}
\circ 
\dot{I}_r \circ \Pi_r \circ \Gamma
+
\Gamma \circ \Pi_r \circ \dot{I}_r
\EC
\end{align}
and, resolving the remaining inverse by the extended BSE \eqref{eq:ebse}, we find
\begin{align}
\dot{\gamma}_r
&
=
\underbrace{
\Gamma \circ \dot{\Pi}_r \circ \Gamma
}_{\dot{\gamma}_r^{(1)}}
+
\underbrace{
\dot{I}_r \circ \Pi_r \circ \Gamma
}_{\dot{\gamma}_r^{\textrm{(L)}}}
+
\underbrace{
\Gamma \circ \Pi_r \circ \dot{I}_r \circ \Pi_r \circ \Gamma
}_{\dot{\gamma}_r^{\textrm{(C)}}}
+
\underbrace{
\Gamma \circ \Pi_r \circ \dot{I}_r
}_{\dot{\gamma}_r^{\textrm{(R)}}}
\ED
\label{eq:gamma_flow}
\end{align}
The algebraic derivation of this exact flow equation, as the differential form of the BSE \eqref{eq:bse}, is our first main result.
It is depicted diagrammatically in Fig.~\ref{fig:gamma_flow}(a) (exemplified by the $a$ channel) and contrasted with the corresponding standard fRG flow equation [Fig.~\ref{fig:gamma_flow}(b)].
It describes the flow of the reducible vertices, $\gamma_r$; the totally irreducible vertex, $R$, does not have an efficient flow equation and remains as input.
Since $\dot{R}=0$, we have $\dot{I}_r = \sum_{r'\neq r} \dot{\gamma}_{r'}  \equiv \dot{\gamma}_{\bar{r}}$,
and Eq.~\eqref{eq:gamma_flow} constitutes a closed, coupled set of differential equations
for all reducible vertices $\gamma_r$.
The natural way to solve these equations
is to start by computing the independent, \textit{one-loop} part,
$\dot{\gamma}_r^{(1)}$, for each channel, and then iteratively
insert the results into the \textit{left}, \textit{right}, and \textit{center} parts [$\dot{\gamma}_r^{\textrm{(L)}}$, $\dot{\gamma}_r^{\textrm{(R)}}$, $\dot{\gamma}_r^{\textrm{(C)}}$, respectively]
of the various channels. If this is organized by the number of 
loops (connecting full vertices), we precisely recover the multiloop fRG (mfRG)
vertex flow which has been derived diagrammatically in Refs.~\onlinecite{Kugler2018,Kugler2018a} (cf.\ Fig.~5 of Ref.~\onlinecite{Kugler2018a}).
It is worth mentioning that the numerical effort of this iterative mfRG flow grows only linearly with the number of loops that are kept (on average) as compared to the standard (truncated) fRG flow \cite{Kugler2018,Kugler2018a,Tagliavini2018}. First implementations \cite{Kugler2018,Tagliavini2018} of this iterative scheme for moderate interaction strengths have found rapid convergence for a number of loops $\lesssim 8$. 
In general, we expect that with increasing interacting strength the convergence with loop order will become slower---and possibly not occur at all for sufficiently strong interactions---in a way that will depend on the model at hand.
From the above derivation, it is clear that,
if the scale dependence of $G$ is chosen such that we are initially able to solve the BSEs (using $G^{\Lambda_i}$) and finally revert to the original theory ($G^{\Lambda_f}=G$), 
then solving the mfRG vertex flow \eqref{eq:gamma_flow} is equivalent to solving the BSEs \eqref{eq:bse}.
An initial solution is always available by using $G^{\Lambda_i}=0$, but can also be chosen differently, if desired (see below).
In the same way that any solution of the BSEs depends on a certain choice of $R$,
so do results of mfRG.
However, the multiloop flow equation requires only the initial condition of the \textit{full} vertex $\Gamma^{\Lambda_i}=R+\sum_r \gamma_r^{\Lambda_i}$ and not of the individual two-particle-reducible or -irreducible vertices;
the decomposition into $\dot{\gamma}_r$ is only performed on the differential level.
Nevertheless, the degree of approximation in our approach is encoded in the underlying expression for $R$,
which can range from the simplest approximation, $R=\Gamma_0$, to the exact object, $R^{\textrm{ex}}$.
\subsubsection{Examples}
\label{sec:vertex_flow_examples}
Let us give some examples for possible flows which are specified by the input $R$ and the choice of $G^{\Lambda_i}$ initializing the progression towards $G^{\Lambda_f}=G$.
Recall that, in this section, we focus on the two-particle level, i.e., we study the influence of varying the full propagator, $G^{\Lambda}$, on the vertex, $\Gamma^{\Lambda}$. In practice, the variation of $G^{\Lambda}$ will be realized by tuning the bare propagator, $G_0^{\Lambda}$, and complementing the vertex flow with a self-energy flow to compute $G^{\Lambda}$ (see Section \ref{sec:self-energy}).
\begin{enumerate}[(i)]
\item
The BSEs at the initial scale are trivially solved if $G^{\Lambda_i}=0$:
Due to $\Pi_r^{\Lambda_i}=0$, the corresponding initial condition for the reducible vertices is $\gamma_r^{\Lambda_i}=0$. As we introduce the scale dependence only for the propagators connecting the vertices in the BSEs but leave the totally irreducible vertex $R$---the input to the parquet equations---unchanged, the initial condition for the full vertex is given by $\Gamma^{\Lambda_i}=R$ \footnote{Whereas the initial condition $\Gamma^{\Lambda_i}=R$ at $G^{\Lambda_i}=0$ is natural in the parquet approximation $R=\Gamma_0$, it might seem counter-intuitive for other cases, when thinking of the totally irreducible vertex, $R$, being itself composed of diagrams containing propagators. In this way of thinking, we have to treat propagators in $R$ differently from those propagators that connect the building block $R$ in the two-particle-reducible diagrams of the $\gamma_r$. This special treatment is necessary as $R$ does not have an efficient flow equation.}.
Hence, the mfRG flow generates all two-particle-reducible diagrams given the irreducible building block $R$;
the special case of $R=\Gamma_0$ yields all diagrams of the parquet approximation (PA) \cite{Kugler2018,Kugler2018a}.
\item 
The mfRG flow \eqref{eq:gamma_flow} is an \textit{exact} flow equation for the two-particle-reducible vertices and thus gives us full control over the vertices corresponding to given propagators $G^{\Lambda}$. Immediate consequences are that (a) for given boundary conditions $G^{\Lambda_i}$, $G^{\Lambda_f}$, we are completely free to choose any \textit{specific} $\Lambda$ dependence in $G^{\Lambda}$---the results of the flow do not depend on this choice; and (b) that we can perform loops in theory space, going from $G^{\Lambda_i}$ to $G^{\Lambda_f}=G^{\Lambda_i}$ without any loss of information. Conceptually, this underlines the power of the mfRG flow; practically, it can also be used as a consistency check for a numerical implementation (which might employ approximate parametrizations of the vertex functions, etc.).
We emphasize that, while both properties directly follow from the given derivation based on the BSEs, they are violated in the widely used one-loop form ($\dot{\gamma}_r \approx \dot{\gamma}_r^{(1)}$) of the truncated fRG flow.

A loop in theory space could for instance be realized via $G^{\Lambda} = f(\Lambda) G^{\Lambda_i}$ with $f(\Lambda_i)=f(\Lambda_f)=1$.
If we already have the result of the parquet approximation ($R=\Gamma_0$) in the form of $G^{\Lambda_i}=G^{\textrm{PA}}$ and $\Gamma^{\Lambda_i}=\Gamma^{\textrm{PA}}$, the vertex flow naturally gives the corresponding parquet vertex for all values of $\Lambda$ (as $R=\Gamma_0$ throughout) and finally returns to the original result.
If we assume (from a conceptual point of view) we had the exact solution of the many-body problem in the form of $G^{\Lambda_i}=G^{\textrm{ex}}$, $\Gamma^{\Lambda_i}=\Gamma^{\textrm{ex}}$, then such a vertex flow would return to the exact result, too. 
However, as the totally irreducible vertex remains fixed, the results at intermediate $\Lambda$ do not correspond to the exact solution for that $G^{\Lambda}$. Instead, at each value of $\Lambda$, the reducible vertices $\gamma_r^{\Lambda}$ solve the BSEs with propagators $G^{\Lambda}$ and $R^{\textrm{ex}} \neq R^{\Lambda}$. At $\Lambda_f$, the BSEs with $G^{\Lambda_f}=G^{\textrm{ex}}$ and $R^{\textrm{ex}}$ reproduce $\gamma_r^{\textrm{ex}}$ and thus $\Gamma^{\textrm{ex}} = R^{\textrm{ex}} + \sum_r \gamma_r^{\textrm{ex}}$.
\item
As a highly correlated and, yet, numerically tractable initial condition \cite{Wentzell2015}, one can choose the solution of dynamical mean-field theory (DMFT) \cite{Georges1996} and use the mfRG flow to generate nonlocal correlations \cite{Taranto2014,Rohringer2017}, thus extending the DMF$^2$RG idea \cite{Taranto2014} to \textit{multiloop} DMF$^2$RG \cite{Kugler2018,Kugler2018a} (or D(MF)$^2$RG \cite{Tagliavini2018a}).
A related approach that gives diagrammatic, nonlocal corrections to DMFT 
is given by the dynamical vertex approximation (D$\Gamma$A) \cite{Toschi2007,Held2008,Valli2015}. 
This approach directly employs the parquet equations, using as input $R^{\textrm{DMFT}}$,
the totally irreducible vertex from the local DMFT solution \cite{Rohringer2012}.
If we used the same initial propagator $G^{\Lambda_i}=0$ as in example (i) above, we would start the vertex flow from $\Gamma^{\Lambda_i}=R^{\textrm{DMFT}}$, in perfect analogy to the D$\Gamma$A algorithm.
However, at this point we can leverage the flexibility of the RG framework and perform a continuous deformation starting directly from the full DMFT vertex: indeed, if we use $G^{\Lambda_i} = G^{\textrm{DMFT}}$ (as opposed to $G^{\Lambda_i}=0$), the vertex flow is not initiated by $R^{\textrm{DMFT}}$, but from the actual, full vertex $\Gamma^{\textrm{DMFT}}$ \cite{Taranto2014}.
(Recall that the decomposition into two-particle channels in Eq.\ \eqref{eq:gamma_flow} occurs only for \textit{differentiated} vertices $\dot{\gamma}_r$, which are ultimately combined to give $\dot{\Gamma} = \sum_r \dot{\gamma}_r$.)
Although the results are (in principle) independent of the specific $\Lambda$ dependence, the choice $G^{\Lambda_i} = G^{\textrm{DMFT}}$ with $\Gamma^{\Lambda_i}=\Gamma^{\textrm{DMFT}}$ has the decisive numerical advantage that it avoids any explicit appearance of $R^{\textrm{DMFT}}$.
The corresponding multiloop flow is hence not affected by the (likely) unphysical divergences of the totally irreducible vertex, which have been observed in strongly correlated systems \cite{Schaefer2013,Schaefer2016,Gunnarsson2017,Chalupa2018,Thunstroem2018}, and can thus be used to analyze such systems in wider regimes of the phase diagram.
The combination of vertex and self-energy flow in multiloop DMF$^2$RG, as used in practice, is further discussed in Section \ref{sec:self-energy_interpretation} (iv).
\end{enumerate}
So far, we have assumed the dressed propagator, $G$, to be known. 
However, as this is in general not the case, we now combine Eq.~\eqref{eq:gamma_flow} with a self-energy flow, $\dot{\Sigma}^{\Lambda}$, to generate $G^{\Lambda}$ during the flow. Via the Dyson equation, we then have $(G^{\Lambda})^{-1} = (G_0^{\Lambda})^{-1} - \Sigma^{\Lambda}$ in a flow controlled by the scale-dependent bare propagator, $G_0^{\Lambda}$.
\section{Derivation of the self-energy flow}
\label{sec:self-energy}
First, let us mention that the straightforward derivation of the vertex flow was based on the parquet equations (for given input $R$). These merely represent a classification of diagrams, reducing the need for an explicit input expression to the most fundamental building block. We did not use equations which provide a construction of the four-point vertex from higher-point vertices, such as the SDE involving $\Gamma^{(6)}$, or a functional derivative connecting four- and six-point vertices.
By contrast, we next want to construct the self-energy, $\Sigma$, from the four-point vertex, $\Gamma$. For this purpose, three equations are available: (i) the SDE relating $\Sigma$ to $\Gamma$, typically used in the parquet formalism \cite{Bickers2004}, (ii) a functional derivative between self-energy and two-particle-irreducible vertex, known from Hedin's equations \cite{Hedin1965} and $\Phi$-derivable approaches \cite{Luttinger1960,Baym1962}, and (iii) the fRG flow equation for $\Sigma$ \cite{Metzner2012}. 
While all these equations are exact, their outcomes might differ when inserting an approximate vertex.
In Section \ref{sec:sigma_funcderiv}, we show that the
fRG flow for $\Sigma$ can be easily derived from the functional derivative (as a necessary condition).
However, as we show in Appendix \ref{sec:sigmaproof}, the SDE and the functional derivative are complementary in the sense that any solution that fulfills both equations must be the exact solution.
It is therefore not surprising that it is complicated to relate a self-energy flow to the SDE for $\Sigma$. Nevertheless, we will use the SDE to derive a self-energy flow (different from the standard fRG flow), which is well-suited for the parquet approximation (PA) and allows us to gain insight into its conservation properties (see Section \ref{sec:conservation}). While this multiloop flow deduced from the SDE indeed proves beneficial in the PA \cite{Kugler2018a}, the general advantages and disadvantages of the different starting points (i) and (ii) are not entirely clear (see also Section \ref{sec:self-energy_interpretation}).
\subsection{Self-energy flow from the Schwinger-Dyson equation}
\label{sec:sde}
\begin{figure}[t]
\includegraphics[width=.84\textwidth]{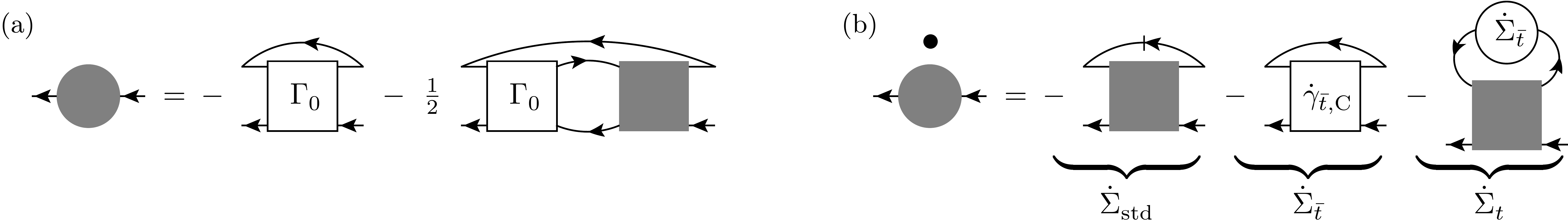}
\caption{%
(a)
Schwinger-Dyson equation (SDE) for the self-energy, where the second term contains two equivalent lines connected to antisymmetric vertices and hence requires a factor of 1/2. One notes that the three propagators in the second summand can be both viewed as contracting a parallel and antiparallel bubble of the vertices $\Gamma_0$ and $\Gamma$.
(b)
Multiloop fRG self-energy flow \cite{Kugler2018a}, derived from the SDE in the parquet approximation.
The first term, $\dot{\Sigma}_{\textrm{std}}$, constitutes the standard fRG self-energy flow.%
}
\label{fig:sigma_sd_flow}
\end{figure}

Deriving a flow equation from the SDE of the self-energy is a difficult task since (as already mentioned) SDEs and differential equations are of fundamentally different nature---for instance, SDEs always contain the bare interaction whereas differential equations are typically phrased with renormalized objects only. In Ref.~\onlinecite{Veschgini2013}, the SDE was used to derive the fRG self-energy flow up to terms $\mathit{O}\big[(\Gamma)^3\big]$; here, we demonstrate agreement up to $\mathit{O}\big[(\Gamma)^4\big]$. In fact, we derive the mfRG self-energy flow from Ref.\ \onlinecite{Kugler2018a}, which includes important terms that would be neglected if one simply inserts the approximate parquet vertex into the standard fRG self-energy flow equation \cite{Kugler2018a}. The calculation with the main results given in Eqs.~\eqref{eq:sigma_dot_g_dot} and \eqref{eq:sigma_flow} (see also Fig.~\ref{fig:sigma_sd_flow}) is presented in detail in the following Section \ref{sec:self-energy_flow} and interpreted in Section \ref{sec:self-energy_interpretation}.
\subsubsection{Flow equation}
\label{sec:self-energy_flow}
\begin{figure}[t]
\includegraphics[width=.78\textwidth]{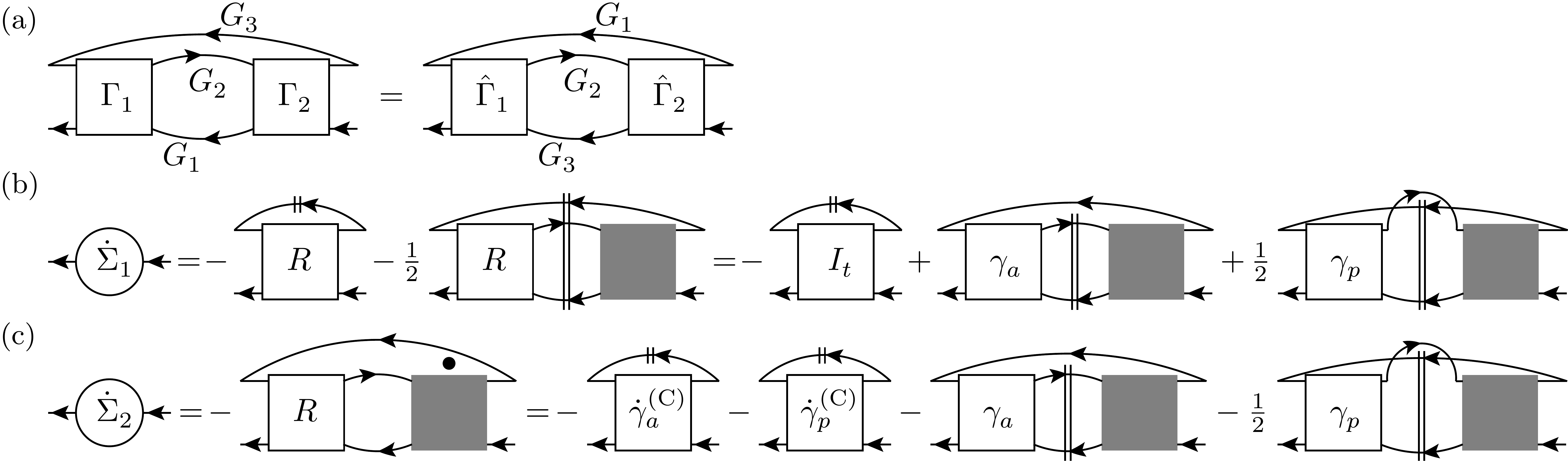}
\caption{%
Illustrations for the derivation of the self-energy flow.
(a)
As stated in Eq.\ \eqref{eq:contracted_crossing},
a bubble of vertices closed with an additional loop can be viewed as a contracted $a$ or $p$ bubble
and can be rewritten by
exchanging two of the external legs ($\Gamma \to \hat{\Gamma}$) of the vertices
(contracted crossing symmetry).
Note that Eq.\ \eqref{eq:contracted_crossing_1} is fully encoded in the diagram on the l.h.s.\ and Eq.\ \eqref{eq:contracted_crossing_2} in the one on the r.h.s. Thus, the individual equations  \eqref{eq:contracted_crossing_1} and  \eqref{eq:contracted_crossing_2} merely represent a redundancy in the algebraic description.
(b)
Rewriting of $\dot{\Sigma}_1$, the part of $\partial_{\Lambda} \Sigma_{\textrm{SD}}$ where the derivative is applied to the propagators appearing explicitly in the SDE. The double dash crossing multiple lines denotes the derivative of the product of propagators, i.e., a sum of terms where each line is differentiated once.
(c)
Rewriting of $\dot{\Sigma}_2$, the part of $\partial_{\Lambda} \Sigma_{\textrm{SD}}$ containing $\dot{\Gamma}$.%
}
\label{fig:contracted_crossing}
\end{figure}

The starting point of our calculation is the \textit{Schwinger-Dyson equation} for the self-energy [cf.\ Fig.~\ref{fig:sigma_sd_flow}(a)]:
\begin{align}
\Sigma = \Sigma_{\textrm{SD}}(\Gamma_0,\Gamma,G)
= - ( \Gamma_0 + \Gamma_0 \circ \Pi_p \circ \Gamma ) \cdot G
= - ( \Gamma_0 + \tfrac{1}{2} \Gamma_0 \circ \Pi_a \circ \Gamma) \cdot G
\ED
\label{eq:sde}
\end{align}
Here, we have used bubbles in either the $a$ or the $p$ channel, as well as the contraction of two vertex legs with a propagator [denoted by $\Gamma \cdot G$, cf.\ Appendix \ref{sec:matrixnotation}, Eq.\ \eqref{eq:self-energy_loop}]. As we can freely choose the specific propagator for the final contraction, we can write the SDE with a bubble in either the $p$ or the $a$ channel---the factor of 1/2 is implicitly contained in $\Pi_p$ and must be explicitly written when using $\Pi_a$.
The presence of two \textit{equivalent} lines [i.e., parallel lines connected to (anti)symmetric vertices] in the second summand of the SDE opens the possibility for further manipulations. For this, let us explicitly denote the propagators contained in a bubble by $\Pi_{r;G_1,G_2}$; the standard bubble is then simply given by $\Pi_r \equiv \Pi_{r;G,G}$. In the SDE, we can not only freely choose the propagator used in the final contraction [Eq.~\eqref{eq:contracted_crossing_1}], we can also switch the equivalent lines by crossing two external legs of both vertices, $\Gamma_1 \to \hat{\Gamma}_1$, $\Gamma_2 \to \hat{\Gamma}_2$ [cf.\ Eq.~\eqref{eq:exchange_legs}]. 
The relations deduced from this \textit{contracted crossing} operation [cf.\ Fig.~\ref{fig:contracted_crossing}(a)] are
\begin{subequations}
\begin{align}
( \tfrac{1}{2} \Gamma_1 \circ \Pi_{a;G_1,G_2} \circ \Gamma_2 ) \cdot G_3
& = ( \Gamma_1 \circ \Pi_{p;G_1,G_3} \circ \Gamma_2 ) \cdot G_2
\label{eq:contracted_crossing_1}
\\
=
( \tfrac{1}{2} \hat{\Gamma}_1 \circ \Pi_{a;G_3,G_2} \circ \hat{\Gamma}_2 ) \cdot G_1
& = ( \hat{\Gamma}_1 \circ \Pi_{p;G_3,G_1} \circ \hat{\Gamma}_2 ) \cdot G_2
\ED
\label{eq:contracted_crossing_2}
\end{align}
\label{eq:contracted_crossing}
\end{subequations}
We will use the contracted crossing relations extensively on the relevant vertices, which obey the crossing symmetries
\begin{equation}
\hat{\Gamma} = -\Gamma \EC \quad
\hat{\Gamma}_0 = -\Gamma_0 \EC \quad
\hat{R} = -R \EC \quad
\hat{\gamma}_p = -\gamma_p \EC \quad
\hat{\gamma}_a = -\gamma_t \EC \quad
\hat{\gamma}_t = -\gamma_a \ED
\label{eq:crossing}
\end{equation}
Note that the vertices in the particle-hole channels $a$, $t$ are mapped onto each other upon crossing two external legs. For this reason, we will often combine contributions from the $a$ and $t$ channel in the following calculations.
The SDE yields a scale-dependent self-energy if we attach a $\Lambda$ dependence to every propagator connecting the vertices in Eq.~\eqref{eq:sde} and account for the $\Lambda$ dependence of the four-point vertex, $\Gamma$, as discussed in Section \ref{sec:vertex}.
In light of the functional derivative $\delta \Sigma / \delta G = -I_t$ (see Section \ref{sec:sigma_funcderiv} below), we aim at generating the irreducible vertex $I_t$, for which we need the totally irreducible vertex, $R$, instead of the bare vertex, $\Gamma_0$. Hence, we define $R'=R-\Gamma_0$, and, since Eq.~\eqref{eq:sde} is linear in $\Gamma_0$, we obtain
\begin{equation}
\Sigma = \Sigma_{\textrm{SD}}(R,\Gamma,G) - \Sigma_{\textrm{SD}}(R',\Gamma,G)
\ED
\label{eq:SDE_Gamma0_R}
\end{equation}
We now consider the flow of $\Sigma_{\textrm{SD}}(R,\Gamma,G)$ and organize our computation according to [cf.\ Figs.~\ref{fig:contracted_crossing}(b) and \ref{fig:contracted_crossing}(c)]
\begin{equation}
\dot{\Sigma} = 
\underbrace{
\partial_{\Lambda}\Sigma_{\textrm{SD}}(R,\Gamma,G) 
- [-(R \circ \Pi_p \circ \dot{\Gamma}) \cdot G]
}_{\dot{\Sigma}_1}
+
\underbrace{
[-(R \circ \Pi_p \circ \dot{\Gamma}) \cdot G]
}_{\dot{\Sigma}_2}
-
\underbrace{
\partial_{\Lambda}\Sigma_{\textrm{SD}}(R',\Gamma,G)
}_{\dot{\Sigma}_3}
\ED
\label{eq:sigma_flow_parts}
\end{equation}
Here, we have subtracted and added a term such that the first bracket, $\dot{\Sigma}_1$, contains only 
those terms of the differentiated SDE in which the derivative is explicitly applied to propagators.
The second part, $\dot{\Sigma}_2$, accounts for the differentiated vertex for which we will insert the vertex flow \eqref{eq:gamma_flow}.
Finally, $\dot{\Sigma}_3$ contains all remaining contributions proportional to $R'$. In the PA, one has $R=\Gamma_0 \Leftrightarrow R'=0$; thus, $\dot{\Sigma}_3$ will only be relevant in calculations that go beyond the PA. In fact, from Eq.~\eqref{eq:SDE_Gamma0_R}, we see that the role of $\dot{\Sigma}_3$ is to cancel the extra terms that have been added to $\dot{\Sigma}_1+\dot{\Sigma}_2$ by using $\Sigma_{\textrm{SD}}(R,\Gamma,G)$ instead of $\Sigma_{\textrm{SD}}(\Gamma_0,\Gamma,G)$.
We begin our calculations with $\dot{\Sigma}_1$.
\textit{Generate $I_t \cdot \dot{G}$}---%
As already mentioned, we want to single out the two-particle-irreducible vertex $I_t$ (since it constitutes the functional derivative of the self-energy). The first summand in Eq.~\eqref{eq:sde} (using $R$ instead of $\Gamma_0$ with $\dot{R}=0$) is easily differentiated as $-R \cdot \dot{G}$. In the remaining part of $\dot{\Sigma}_1$, we have three propagators to differentiate. Two of the resulting terms can be combined to factor out $\dot{G}$ if we use the contracted crossing symmetry \eqref{eq:contracted_crossing} on $R$ and $\Gamma$:
\begin{align}
-\dot{\Sigma}_1 
-
R \cdot \dot{G}
& = 
\big( R \circ ( \Pi_{p;\dot{G},G} + \Pi_{p;G,\dot{G}} ) \circ \Gamma \big) \cdot G
+
( R \circ \Pi_{p;G,G} \circ \Gamma ) \cdot \dot{G}
=
( R \circ \Pi_{a} \circ \Gamma ) \cdot \dot{G}
+
( R \circ \Pi_{p} \circ \Gamma ) \cdot \dot{G}
\ED
\end{align}
Next, we collect the terms for 
$I_t = R + \gamma_{\bar{t}} = R + I_a \circ \Pi_a \circ \Gamma + I_p \circ \Pi_p \circ \Gamma$ 
[cf.\ Eq.~\eqref{eq:parquet}]
and find
\begin{align}
-\dot{\Sigma}_1 
& = 
\big[
R
+ 
( R \circ \Pi_{a} \circ \Gamma )
+
( R \circ \Pi_{p} \circ \Gamma )
\big]
\cdot \dot{G}
=
I_t \cdot \dot{G} - 
\big[
(\gamma_p+\gamma_t) \circ \Pi_a \circ \Gamma
+ 
(\gamma_a+\gamma_t) \circ \Pi_p \circ \Gamma)
\big] \cdot \dot{G}
\ED
\label{eq:It-extra_terms}
\end{align}
\textit{Use differentiated bubbles}---%
The extra terms accompanying $I_t \cdot \dot{G}$ in Eq.\ \eqref{eq:It-extra_terms} will later be combined with contributions from $\dot{\Sigma}_2$. 
Since $\dot{\Sigma}_2$ contains the differentiated vertex, which itself is built from differentiated bubbles $\dot{\Pi}_r$, we rewrite these contributions in terms of $\dot{\Pi}_r$. Using the contracted crossing symmetry \eqref{eq:contracted_crossing}, we find
\begin{subequations}
\begin{align}
( \gamma_p \circ \Pi_{a;G,G} \circ \Gamma ) \cdot \dot{G}
& =
( \gamma_p \circ \Pi_{p;G,\dot{G}} \circ \Gamma ) \cdot G
+
( \gamma_p \circ \Pi_{p;\dot{G},G} \circ \Gamma ) \cdot G
=
( \gamma_p \circ \dot{\Pi}_p \circ \Gamma ) \cdot G
\EC
\\
( \gamma_t \circ \Pi_{a;G,G} \circ \Gamma ) \cdot \dot{G}
& =
( \gamma_a \circ \Pi_{a;\dot{G},G} \circ \Gamma ) \cdot G
\EC
\\
\big[ ( \gamma_a + \gamma_t ) \circ \Pi_{p;G,G} \circ \Gamma \big] \cdot \dot{G}
& =
( 2 \gamma_a \circ \Pi_{p;G,G} \circ \Gamma ) \cdot \dot{G}
=
( \gamma_a \circ \Pi_{a;G,\dot{G}} \circ \Gamma ) \cdot G
\ED
\end{align}
\end{subequations}
This leads to the final expression for $\dot{\Sigma}_1$ [illustrated in Fig.~\ref{fig:contracted_crossing}(b)]:
\begin{align}
\dot{\Sigma}_1 
& =
-I_t \cdot \dot{G} +
( \gamma_a \circ \dot{\Pi}_a \circ \Gamma + \gamma_p \circ \dot{\Pi}_p \circ \Gamma ) \cdot G
\ED
\label{eq:sigma_dot_1}
\end{align}
\textit{Organize vertex derivative}---%
The second contribution to Eq.\ \eqref{eq:sigma_flow_parts}, $\dot{\Sigma}_2$, contains the differentiated vertex.
Inserting the decomposition $\dot{\Gamma} = \sum_r \dot{\gamma}_r$, we can combine the contributions from both particle-hole channels, $a$ and $t$, by applying the contracted crossing symmetry \eqref{eq:contracted_crossing} on $R$ and $\dot{\gamma}_t$:
\begin{equation}
-\dot{\Sigma}_2 = (R \circ \Pi_p \circ \dot{\Gamma}) \cdot G
= (R \circ \Pi_a \circ \dot{\gamma}_a) \cdot G + (R \circ \Pi_p \circ \dot{\gamma_p}) \cdot G
\ED
\label{eq:sigma_dot_2}
\end{equation}
Once we insert the flow equation \eqref{eq:gamma_flow} for $\dot{\gamma}_a$ and $\dot{\gamma}_p$ 
in Eq.\ \eqref{eq:sigma_dot_2}, $R$ will be connected to further bubbles of vertices.
These connections can be simplified if we have $I_r$ instead of $R$. Hence, we rewrite Eq.\ \eqref{eq:sigma_dot_2},
using $I_r = R + \gamma_{\bar{r}}$, as
\begin{equation}
-\dot{\Sigma}_2 =
(I_a \circ \Pi_a \circ \dot{\gamma}_a) \cdot G 
- [(\gamma_p+\gamma_t) \circ \Pi_a \circ \dot{\gamma}_a] \cdot G 
+ (I_p \circ \Pi_p \circ \dot{\gamma_p}) \cdot G
- [(\gamma_a+\gamma_t) \circ \Pi_p \circ \dot{\gamma}_p] \cdot G 
\ED
\end{equation}
The next step consists of repeated use of the contracted crossing symmetry \eqref{eq:contracted_crossing}:
\begin{subequations}
\begin{align}
(\gamma_p \circ \Pi_a \circ \dot{\gamma}_a) \cdot G
& =
(\gamma_p \circ \Pi_p \circ \dot{\gamma}_a) \cdot G 
+ (\gamma_p \circ \Pi_p \circ \dot{\gamma}_t) \cdot G 
\EC \\
(\gamma_t \circ \Pi_a \circ \dot{\gamma}_a) \cdot G
& =
(\gamma_a \circ \Pi_a \circ \dot{\gamma}_t) \cdot G 
\EC \\
[(\gamma_a+\gamma_t) \circ \Pi_p \circ \dot{\gamma}_p] \cdot G
& =
(\gamma_a \circ \Pi_a \circ \dot{\gamma}_p) \cdot G 
\ED
\end{align}
\end{subequations}
After using $\dot{I}_r = \dot{\gamma}_{\bar{r}}$, we then obtain
\begin{align}
\dot{\Sigma}_2 
& = -
\sum_{r=a,p}
\big(
I_r \circ \Pi_r \circ \dot{\gamma}_r
-
\gamma_r \circ \Pi_r \circ \dot{I}_r
\big)
\cdot G 
\ED
\label{eq:sigma_dot_2_sum}
\end{align}
\textit{Insert vertex flow}---%
Whereas the previous manipulations were possible due to the contracted crossing symmetry,
the following insertion of the vertex flow for $\dot{\gamma}_r$, given by Eq.~\eqref{eq:gamma_flow}, can be simplified already on the vertex level.
In fact, using the parquet equations \eqref{eq:parquet} with
$\gamma_r = I_r \circ \Pi_r \circ \Gamma$ and $\Gamma = I_r + \gamma_r$, we get
\begin{align}
I_r \circ \Pi_r \circ \dot{\gamma}_r
& =
I_r \circ \Pi_r \circ
\big( \Gamma \circ \dot{\Pi}_r \circ \Gamma + 
\dot{I}_r \circ \Pi_r \circ \Gamma +
\Gamma \circ \Pi_r \circ \dot{I}_r \circ \Pi_r \circ \Gamma +
\Gamma \circ \Pi_r \circ \dot{I}_r \big)
\nonumber \\
& =
\gamma_r \circ \dot{\Pi}_r \circ \Gamma + 
\Gamma \circ \Pi_r \circ \dot{I}_r \circ \Pi_r \circ \Gamma +
\gamma_r \circ \Pi_r \circ \dot{I}_r
\ED
\label{eq:sigma_dot_2_part}
\end{align}
The first term also occurs (with opposite sign) in Eq.~\eqref{eq:sigma_dot_1},
the second term reproduces $\dot{\gamma}_r^{(\textrm{C})}$,
and the third term gets canceled in Eq.~\eqref{eq:sigma_dot_2_sum}. 
Hence, $\dot{\Sigma}_2$ can be simplified [as summarized in Fig.~\ref{fig:contracted_crossing}(c)] to
\begin{equation}
\dot{\Sigma}_2 = -\sum_{r=a,p} \big( \dot{\gamma}_r^{(\textrm{C})} + \gamma_r \circ \dot{\Pi}_r \circ \Gamma \big) \cdot G
\ED
\label{eq:sigma_dot_2_final}
\end{equation}
With the definition 
$\dot{\gamma}_{\bar{t}}^{\textrm{(C)}} =  \dot{\gamma}_a^{(\textrm{C})} + \dot{\gamma}_p^{(\textrm{C})}$,
the full derivative of the self-energy is given by
\begin{align}
\dot{\Sigma}
& =
\dot{\Sigma}_1 + \dot{\Sigma}_2 - \dot{\Sigma}_3
=
-I_t \cdot \dot{G}
- \dot{\gamma}_{\bar{t}}^{\textrm{(C)}} \cdot G
- \dot{\Sigma}_3
\ED
\label{eq:sigma_dot_g_dot}
\end{align}
This result for $\dot{\Sigma} \equiv \partial_{\Lambda} \Sigma_{\textrm{SD}}$ in \textit{skeleton} form (i.e., phrased with dressed propagators $G$, $\dot{G}$ only) will be considered more closely in Section \ref{sec:conservation}.
Here, we move on by noting that Eq.~\eqref{eq:sigma_dot_g_dot} still contains $\dot{\Sigma}$ on both the l.h.s.\ and the r.h.s.\ (via $\dot{G}$).
\textit{Isolate $\dot{\Sigma}$}---%
At this point in our derivation, we specify how the $\Lambda$ dependence is supposed to enter $G$: it shall be incorporated in the bare propagator $G_0$ such that the Dyson equation, $G=G_0 + G_0 \cdot \Sigma \cdot G$, entails $\dot{G} = S + G \cdot \dot{\Sigma} \cdot G$ with the single-scale propagator $S=\partial_{\Lambda}G_{|_{\Sigma=\textrm{const}}}= - G \cdot (\partial_{\Lambda} G_0^{-1} ) \cdot G$.
Once we insert this expression for $\dot{G}$ into Eq.~\eqref{eq:sigma_dot_g_dot}, we will face the contraction of a vertex with a composite line $G \cdot \dot{\Sigma} \cdot G$. In such a case, one can equivalently attribute the two propagators to either the self-energy or the vertex, such that we have the following equality for a \textit{composite contraction} [recall the minus sign in $\Pi_t$; see Eq.~\eqref{eq:loopbubble} for details]:
\begin{equation}
I_t \cdot (G \cdot \dot{\Sigma} \cdot G)
=
- I_t \circ \Pi_t \cdot \dot{\Sigma}
\ED
\label{eq:comp-contr}
\end{equation}
We insert Eq.~\eqref{eq:comp-contr} into Eq.~\eqref{eq:sigma_dot_g_dot} to isolate $\dot{\Sigma}$:
\begin{align}
\dot{\Sigma}
& 
=
-I_t \cdot (S+ G \cdot \dot{\Sigma} \cdot G)
- \dot{\gamma}_{\bar{t}}^{\textrm{(C)}} \cdot G - \dot{\Sigma}_3
=
-I_t \cdot S
+ I_t \circ \Pi_t \circ \dot{\Sigma}
- \dot{\gamma}_{\bar{t}}^{\textrm{(C)}} \cdot G 
- \dot{\Sigma}_3
\nonumber \\
\Leftrightarrow \quad
\dot{\Sigma}
&
=
-(1-I_t \circ \Pi_t)^{-1} \circ I_t
\cdot S
-
(1-I_t \circ \Pi_t)^{-1} 
\cdot \big( 
\dot{\gamma}_{\bar{t}}^{\textrm{(C)}} \cdot G + \dot{\Sigma}_3 
\big)
\ED
\end{align}
Next, we use the inverted BSE \eqref{eq:ibse} as well as the extended BSE \eqref{eq:ebse} to express this through $\Gamma$ and $1+\Gamma \circ \Pi_t$, respectively:
\begin{align}
\dot{\Sigma}
=
-\Gamma_{\vphantom{\bar{t}}}^{\vphantom{\textrm{(C)}}} 
\cdot S
-
(1 + \Gamma \circ \Pi_t) \cdot
(
\dot{\gamma}_{\bar{t}}^{\textrm{(C)}} \cdot G
+
\dot{\Sigma}_3
)
\ED
\end{align}
For convenience, we finally write the contraction of $(\Gamma \circ \Pi_t)$ with both summands as composite contractions [using Eq.~\eqref{eq:comp-contr} for a general vertex and self-energy] and obtain
\begin{align}
\dot{\Sigma}
=
\underbrace{[-
\Gamma_{\vphantom{\bar{t}}}^{\vphantom{\textrm{(C)}}} 
\cdot S
]}_{\dot{\Sigma}_{\textrm{std}}}
+ 
\underbrace{[-
\dot{\gamma}_{\bar{t}}^{\textrm{(C)}} \cdot G
]}_{\dot{\Sigma}_{\bar{t}}}
+
\underbrace{[-
\Gamma_{\vphantom{\bar{t}}}^{\vphantom{\textrm{(C)}}} 
 \cdot (G \cdot \dot{\Sigma}_{\bar{t}} \cdot G)
]}_{\dot{\Sigma}_{t}}
- \,
\dot{\Sigma}_3
-
[-\Gamma \cdot ( G \cdot \dot{\Sigma}_3 \cdot G)]
\ED
\label{eq:sigma_flow}
\end{align}
This is our final result for the mfRG self-energy flow deduced from the SDE. 
It constitutes the bare (``nonskeleton'') form of Eq.~\eqref{eq:sigma_dot_g_dot} as it involves $G$ and $S$ instead of $G$ and $\dot{G}$.
The first term in Eq.~\eqref{eq:sigma_flow}, $\dot{\Sigma}_{\textrm{std}}$, is the standard fRG self-energy flow.
The next two terms, $\dot{\Sigma}_{\bar{t}}$ and $\dot{\Sigma}_t$, constitute the multiloop corrections to the self-energy flow [cf.\ Fig.~\ref{fig:sigma_sd_flow}(b)], which have been derived diagrammatically in Ref.~\onlinecite{Kugler2018a}.
These contributions are needed to ensure that the self-energy flow generates all contributions to the self-energy arising within the PA.
Finally, the two terms involving $\dot{\Sigma}_3$ remain in our final result and---in calculations beyond the PA---are required to cancel doubly counted terms coming from the replacement $\Sigma_{\textrm{SD}}(\Gamma_0,\Gamma,G) \rightarrow \Sigma_{\textrm{SD}}(R,\Gamma,G)$ in Eq.~\eqref{eq:SDE_Gamma0_R}.
We remark that $\dot{\Sigma}_3$ constitutes precisely the part that cannot be simplified further with our parquet tools, as it originates from the appearance of a \textit{bare} instead of renormalized vertex in the SDE.
\subsubsection{Interpretation}
\label{sec:self-energy_interpretation}
Let us interpret the flow equation \eqref{eq:sigma_flow} step by step:
\begin{enumerate}[(i)]
\item
Since $\dot{\gamma}_{\bar{t}}^{\textrm{(C)}}$ and $R'$ [and hence $\dot{\Sigma}_3 = \partial_{\Lambda}\Sigma_{\textrm{SD}}(R',\Gamma,G)$] are of order $\mathit{O}\big[(\Gamma)^4\big]$,
we have explicitly shown how to derive the standard fRG self-energy flow, $\dot{\Sigma}_{\textrm{std}}$, from the SDE up to and including terms of fourth order in the (effective) interaction.
If we were in the standard fRG setting where \textit{every} line is $\Lambda$-dependent, further terms coming from $\dot{R}\neq 0$ would arise in our derivation. However, as these terms are similarly of order $\mathit{O}\big[(\Gamma)^4\big]$, the result $\partial_{\Lambda} \Sigma_{\textrm{SD}} = \dot{\Sigma}_{\textrm{std}} + \mathit{O}\big[(\Gamma)^4\big]$ would remain unchanged.
\item
In the PA, the totally irreducible vertex is reduced to its simplest approximation, such that $R=\Gamma_0 \Leftrightarrow R'=0$
and thus
$\dot{\Sigma}_3=0$.
In this case, Eq.~\eqref{eq:sigma_flow} reproduces the mfRG self-energy flow from Ref.~\onlinecite{Kugler2018a} including the corrections $\dot{\Sigma}_{\bar{t}}$ and $\dot{\Sigma}_t$ [cf.\ Fig.~\ref{fig:sigma_sd_flow}(b)], necessary to provide a total derivative of the SDE using the approximate parquet vertex.
\item
Let us come back to the idea of a loop in theory space, which---including the self-energy flow---is now driven by the bare propagator $G_0^{\Lambda}$. 
A possible realization is given by $G_0^{\Lambda} = f(\Lambda)G_0$ with $f(\Lambda_i)=f(\Lambda_f)=1$.
If we start the flow from the solution in the PA ($R=\Gamma_0$) with $\Sigma^{\Lambda_i}=\Sigma^{\textrm{PA}}$ and $\Gamma^{\Lambda_i}=\Gamma^{\textrm{PA}}$, the combination of the mfRG vertex flow \eqref{eq:gamma_flow} and self-energy flow \eqref{eq:sigma_flow} (using $\dot{\Sigma}_3=0$) gives the corresponding result in the PA for all $\Lambda$ (as $R=\Gamma_0$ throughout) and returns to the original solution at $\Lambda_f$.
However, starting the flow from a solution with $R'\neq 0$, we would have to include $\dot{\Sigma}_3$ in the self-energy flow \eqref{eq:sigma_flow} in order to precisely return to the original self-energy, $\Sigma$, and vertex, $\Gamma$ (dressed by $\Sigma$), at $\Lambda_f$;
with $R'\neq 0$, setting $\dot{\Sigma}_3=0$ introduces an approximation in the full derivative of the SDE. 
Conversely, one can compare results of the flow at $\Lambda_i$ and $\Lambda_f$ to (numerically) gauge the importance of the individual terms in Eq.~\eqref{eq:sigma_flow}.
To better understand the effect of $\dot{\Sigma}_3$, we recall that
$\dot{\Sigma}_{\bar{t}}$ and $\dot{\Sigma}_t$ were originally derived diagrammatically to compensate for missing diagrams of $\dot{\Sigma}^{\textrm{PA}}$ when using the parquet vertex in $\dot{\Sigma}_{\textrm{std}}$ \cite{Kugler2018a}.
With this perspective on $\dot{\Sigma}_{\bar{t}} + \dot{\Sigma}_t$ in mind, it is intuitively clear that higher-order contributions to $R$ (i.e., $R'\neq 0$) generate doubly counted terms between $\dot{\Sigma}_{\textrm{std}}$ and $\dot{\Sigma}_{\bar{t}}+\dot{\Sigma}_t$. Yet, as Eq.~\eqref{eq:sigma_flow} is exact, these overcounted terms are precisely canceled by the parts involving $\dot{\Sigma}_3$.

For illustration, consider the (parquet) self-energy at fourth order in the interaction, which contains no approximation and whose flow is fully described by $\dot{\Sigma}_{\textrm{std}}+\dot{\Sigma}_{\bar{t}}+\dot{\Sigma}_t$ using vertices in the PA. Now, fourth-order diagrams of $R'\neq 0$ generate fourth-order terms in $\dot{\Sigma}_{\textrm{std}}$ but not in $\dot{\Sigma}_{\bar{t}}$ and $\dot{\Sigma}_t$ (due to their structure involving further vertices that raise the interaction order). The additional fourth-order contributions of $\dot{\Sigma}_{\textrm{std}}$ are precisely canceled by $R' \cdot \dot{G}$ (containing only \textit{one} $\Lambda$-dependent line) as part of $\dot{\Sigma}_3$.
Generally, we believe that, for situations where $R' \neq 0$, the overcounting of differentiated diagrams in $\dot{\Sigma}_{\textrm{std}}+\dot{\Sigma}_{\bar{t}}+\dot{\Sigma}_t$ has rather small weight and that, even if using $\dot{\Sigma}_3 \approx 0$, the multiloop additions $\dot{\Sigma}_{\bar{t}}+\dot{\Sigma}_t$ provide an improvement of the standard self-energy flow, $\dot{\Sigma}_{\textrm{std}}$. 
\item
An interesting application with $R' \neq 0$ is the previously mentioned multiloop DMF$^2$RG approach. In its full form, combining the flow equations of the vertex \eqref{eq:gamma_flow} and self-energy \eqref{eq:sigma_flow}, the mfRG flow is controlled by the bare propagator $G_0^{\Lambda}$, which interpolates between the local theory of DMFT and the actual lattice problem. The simplest realization \cite{Taranto2014} of a flow from $\Lambda_i=1$ to $\Lambda_f=0$, formulated in terms of Matsubara frequencies $i\omega$ and momentum $\vec{k}$, is given by 
$(G_0^{\Lambda})^{-1} = i\omega + \mu - \Lambda \Delta(i\omega) - (1-\Lambda) \epsilon_{\vec{k}}$.
Here, $\Delta(i\omega)$ is the self-consistently determined hybridization function of the auxiliary Anderson impurity model \cite{Georges1996} and $\epsilon_{\vec{k}}$ the lattice dispersion.
With $G_0^{\Lambda_i}=G_0^{\textrm{DMFT}}=1/[i\omega+\mu-\Delta(i\omega)]$, the flow is conveniently started from $\Sigma^{\Lambda_i}=\Sigma^{\textrm{DMFT}}$ and $\Gamma^{\Lambda_i}=\Gamma^{\textrm{DMFT}}$.
While the vertex flow \eqref{eq:gamma_flow} exactly solves the BSEs (for given $G^{\Lambda}$), the differential form of the SDE contains $\dot{\Sigma}_3$ and therefore prevents complete equivalence to the D$\Gamma$A approach. 
In this regard, it remains to be seen whether the standard fRG self-energy flow, $\dot{\Sigma}_{\textrm{std}}$, with or without the multiloop corrections $\dot{\Sigma}_{\bar{t}}+\dot{\Sigma}_{t}$, or other realizations, incorporating parts of $\dot{\Sigma}_3$ in Eq.~\eqref{eq:sigma_flow}, lead to optimal results.
\end{enumerate}
\subsection{Self-energy flow from the functional derivative}
\label{sec:sigma_funcderiv}
We now show how the standard fRG self-energy flow, $\dot{\Sigma}_{\textrm{std}}$, can be directly derived from the equality between the functional derivative of the self-energy and the (particle-hole) two-particle-irreducible vertex.
To be in perfect accordance with the standard fRG setup, we have to require that \textit{every} $G$ line be $\Lambda$-dependent---even those in the totally irreducible vertex, $R=R^{\Lambda}$. Incorporating the $\Lambda$ dependence in the bare propagator $G_0$, we again relate the differentiated propagator, $\dot{G}$, to the single-scale propagator, $S$, via $\dot{G} = S + G \cdot \dot{\Sigma} \cdot G$.
The functional derivative between self-energy and vertex, $\delta \Sigma / \delta G = -I_t$ [cf.\ Eq.~\eqref{eq:FunctionalDerivative}],
holds for any variation of $G$. 
If this variation is realized by having a scale-dependent propagator $G^{\Lambda}$ and varying the scale parameter $\Lambda$, this equation \textit{implies}
$\dot{\Sigma}=-I_t \cdot \dot{G}$.
Starting from this, we can perform the same steps as above:
To obtain the standard fRG flow equation for the self-energy, it remains to insert
$\dot{G} = S + G \cdot \dot{\Sigma} \cdot G$, 
express the composite contraction $I_t \cdot (G \cdot \dot{\Sigma} \cdot G)$ as
$-I_t \circ \Pi_t \cdot \dot{\Sigma}$ [cf.\ Eq.~\eqref{eq:comp-contr}], and use the inverted BSE \eqref{eq:ibse}:
\begin{align}
\dot{\Sigma}
& =
-I_t \cdot \dot{G}
=
-I_t \cdot (S + G \cdot \dot{\Sigma} \cdot G)
=
-I_t \cdot S + I_t \circ \Pi_t \cdot \dot{\Sigma}
\nonumber \\
\Leftrightarrow \quad
\dot{\Sigma}
& 
= - (1 - I_t \circ \Pi_t)^{-1} \circ I_t \cdot S
=
- \Gamma \cdot S
\ED
\label{eq:sigma_dot_std}
\end{align}
Solving for $\Sigma$ in a specific fRG flow via Eq.~\eqref{eq:sigma_dot_std} amounts to integrating $\delta \Sigma = - I_t \cdot \delta G$ along a specific path in the space of theories defined by the bare propagator $G_0 = G_0^{\Lambda}$ [and the bare interaction $\Gamma_0$, cf.\ Eq.~\eqref{eq:action}]. 
Only if this integration is independent of the path, i.e., if $\dot{\Sigma}$ contains a total derivative of diagrams, the standard self-energy flow \eqref{eq:sigma_dot_std} yields results consistent with the functional derivative. 
In the scenarios considered so far, this is not the case:
the truncated fRG flow (without $\Gamma^{(6)}$ and more than one channel) employs Eq.~\eqref{eq:sigma_dot_std} but does not generate a total derivative of diagrams \cite{Kugler2018,Kugler2018a};
the mfRG flow of Fig.~\ref{fig:sigma_sd_flow} with $R=\Gamma_0$ does provide a total derivative of diagrams but deviates from Eq.~\eqref{eq:sigma_dot_std} by the additions $\dot{\Sigma}_{\bar{t}}$ and $\dot{\Sigma}_t$. (In fact, the latter reproduces precisely the self-energy diagrams generated by the SDE using the vertex in the PA. However, as shown in Appendix \ref{sec:sigmaproof}, the requirement of fulfilling both the functional derivative and the SDE necessitates the exact solution.)
As a direct application of the above calculations, we can derive a fRG flow which is equivalent to self-consistent Hartree-Fock (HF), in agreement with a result by Katanin \cite{Katanin2004}. This conserving fRG flow provides a simple example for which the integration of $\delta \Sigma = - I_t \cdot \delta G$ is indeed independent of the path. 
In HF theory, the functional derivative of the self-energy is given by the bare vertex, $\delta \Sigma^{\textrm{HF}} / \delta G = - \Gamma_0$. By replacing $I_t \to \Gamma_0$ in Eq.~\eqref{eq:sigma_dot_std}, we immediately find
\begin{subequations}
\begin{align}
\dot{\Sigma}^{\textrm{HF}}
& =
- \Gamma_0 \cdot \dot{G}
=
- (1 - \Gamma_0 \circ \Pi_t)^{-1} \circ \Gamma_0 \cdot S
= - \Gamma_t^{\textrm{lad}} \cdot S
\EC 
\label{eq:HF_sigma_flow}
\\
\Gamma_t^{\textrm{lad}}
& =
\Gamma_0 + \Gamma_0 \circ \Pi_t \circ \Gamma_t^{\textrm{lad}}
\quad
\Leftrightarrow
\quad
\Gamma_t^{\textrm{lad}}
=
(1 - \Gamma_0 \circ \Pi_t)^{-1} \circ
\Gamma_0
\EC \\
\dot{\Gamma}_t^{\textrm{lad}}
& =
\Gamma_0 \circ \dot{\Pi}_t \circ \Gamma_t^{\textrm{lad}}
+
\Gamma_0 \circ \Pi_t \circ \dot{\Gamma}_t^{\textrm{lad}}
\
\Leftrightarrow
\
\dot{\Gamma}_t^{\textrm{lad}}
=
(1 - \Gamma_0 \circ \Pi_t)^{-1} \circ
\Gamma_0 \circ \dot{\Pi}_t \circ \Gamma_t^{\textrm{lad}}
=
\Gamma_t^{\textrm{lad}} \circ \dot{\Pi}_t \circ \Gamma_t^{\textrm{lad}}
\ED
\label{eq:HF_vertex_flow}
\end{align}
\end{subequations}
Equation \eqref{eq:HF_vertex_flow} describes the vertex flow in the truncated Katanin form
\footnote{%
The substitution $S \to \dot{G}$ in the truncated fRG vertex flow is often called Katanin substitution \cite{Katanin2004}.}, 
restricted to the $t$ channel. If the same vertex is used for the standard self-energy flow [Eq.~\eqref{eq:HF_sigma_flow}], the fRG flow yields the Hartree-Fock self-energy together with a particle-hole ladder vertex (note $\hat{\Gamma}_t^{\textrm{lad}} =  -\Gamma_a^{\textrm{lad}}$). As this vertex consists of ladder diagrams in only one channel, it clearly violates crossing symmetry.
\section{Conservation laws in the parquet approximation}
\label{sec:conservation}
\begin{figure}[t]
\includegraphics[width=0.85\textwidth]{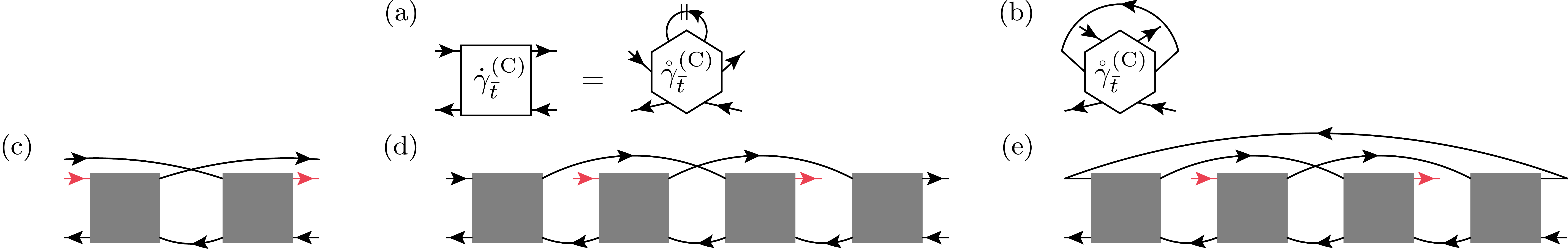} 
\caption{%
Illustrations for $\mathring{\gamma}_r^{(\textrm{C})}$.
(a)
The six-point vertex $\mathring{\gamma}_r^{(\textrm{C})}$ is obtained from $\dot{\gamma}_r^{(\textrm{C})}$ by removing its differentiated line; hence, $\dot{\gamma}_r^{(\textrm{C})}$ is recovered by contracting $\mathring{\gamma}_r^{(\textrm{C})}$ with $\dot{G}$.
(b)
A contraction of $\mathring{\gamma}_r^{(\textrm{C})}$ denoted as $\mathring{\gamma}_r^{(\textrm{C})} \cdot G$, such that $\dot{\gamma}_r^{(\textrm{C})} \cdot G$ is reproduced by $(\mathring{\gamma}_r^{(\textrm{C})} \cdot G) \cdot \dot{G}$.
(c)
As an example for the construction of $\mathring{\gamma}_r^{(\textrm{C})}$, we consider the
six-point vertex $\mathring{\gamma}_p^{(1)}$ obtained by removing the differentiated line in the \textit{one-loop} part of the vertex flow in $p$ channel, leaving two further amputated legs (marked in light red).
(d)
Inserting the vertex from (c) into the \textit{center} part of the flow in the $a$ channel, we generate a contribution to the six-point vertex $\mathring{\gamma}_a^{(\textrm{C})}$ (being part of $\mathring{\gamma}_{\bar{t}}^{(\textrm{C})}$).
(e)
By contracting two upper legs of the vertex from (d) according to $\mathring{\gamma}_{\bar{t}}^{\textrm{(C)}} \cdot G$, we get a contribution to the new, two-particle-irreducible vertex $I_t'$. The lowest-order realization of this, obtained by inserting a bare vertex for each $\Gamma$, constitutes an envelope diagram, which is not contained in the initial $I_t$ in the PA.%
}
\label{fig:bubble_cut}
\end{figure}
In this section, we take a slightly different perspective and are not concerned with RG flows.
Instead, we use our insight into the structure of the many-body relations gained from the above derivations to address conceptual questions of many-body (parquet) theory.
First, we derive two technical results:
(i) We show how one can \textit{construct} a two-particle-irreducible vertex which equals the functional derivative of the parquet self-energy. Evidently, the operation $\delta \Sigma / \delta G$ can be performed in an analytical study of Feynman diagrams \cite{Kugler2018c}. However, in a numerical treatment, one never has access to the self-energy as a \textit{functional} of the full propagator. Instead, one only has its value for the specific, given propagator, and the general construction for such a vertex remains unknown \cite{Smith1992}. Here, we provide its construction for the case of the parquet self-energy.
(ii) We demonstrate that the parquet self-energy can be obtained from the SDE using either of two possible orderings of the bare and full vertex. While it is believed that most approximations for $\Sigma$ obtained from the SDE obey this property \cite{Baym1961}, it has (to our knowledge) not been shown for the PA.
These results can then be interpreted in the context of conservation laws in the PA using arguments from Baym and Kadanoff \cite{Baym1961}.

\subsection{Functional derivative of the parquet self-energy}
We start from the flow equation for the self-energy in skeleton form: In the PA, we have $R=\Gamma_0$, and thus $R'=0$ and $\dot{\Sigma}_3=0$, such that Eq.~\eqref{eq:sigma_dot_g_dot} reads
\begin{align}
\dot{\Sigma}^{\textrm{PA}}
& =
- I_t \cdot \dot{G}
- \dot{\gamma}_{\bar{t}}^{\textrm{(C)}} \cdot G
\ED
\label{eq:sigma_dot_PA}
\end{align}
As $R$ is here given by the bare vertex, our construction of a scale-dependent $\Gamma$ (Section \ref{sec:vertex}) and $\Sigma$ (Section \ref{sec:self-energy}) actually makes \textit{every} propagator scale-dependent. Furthermore, this scale dependence is completely arbitrary, and we can view the scale derivative of the self-energy as coming from the chain rule, $\dot{\Sigma} = (\delta \Sigma / \delta G) \cdot \dot{G}$.
Regarding Eq.~\eqref{eq:sigma_dot_PA}, we want to similarly factorize $\dot{G}$ from the term $\dot{\gamma}_{\bar{t}}^{\textrm{(C)}} \cdot G$. For this, let
$\mathring{\gamma}_{\bar{t}}^{\textrm{(C)}}$ be the six-point vertex obtained from 
$\dot{\gamma}_{\bar{t}}^{\textrm{(C)}}$ be removing the differentiated line, such that $\dot{\gamma}_{\bar{t}}^{\textrm{(C)}}$ is recovered by a contraction with $\dot{G}$, and
$\dot{\gamma}_{\bar{t}}^{\textrm{(C)}} \cdot G = (\mathring{\gamma}_{\bar{t}}^{\textrm{(C)}} \cdot G) \cdot \dot{G}$ [cf.\ Figs.~\ref{fig:bubble_cut} (a) and \ref{fig:bubble_cut}(b)].
It then follows from Eq.~\eqref{eq:sigma_dot_PA} that
\begin{align}
\frac{\delta \Sigma^{\textrm{PA}}}{\delta G}
& =
- I_t 
- \mathring{\gamma}_{\bar{t}}^{\textrm{(C)}} \cdot G
\equiv 
- I_t'
\ED
\label{eq:dsigma_PA}
\end{align}
Here, $\Sigma^{\textrm{PA}}$ is the self-energy obtained from the SDE in the PA (using the vertex $\Gamma = I_t + I_t \circ \Pi_t \circ \Gamma$), and $I_t'$ is the (new) two-particle-irreducible vertex that results from a functional derivative of the parquet self-energy. (The corresponding full vertex $\Gamma'$ can be obtained by solving $\Gamma' = I_t' + I_t' \circ \Pi_t \circ \Gamma'$.)
The crucial point is that---instead of taking the functional derivative---we can \textit{construct} this vertex $I_t'$ by taking the (initial) vertex $I_t$ in the PA and adding the term $\mathring{\gamma}_{\bar{t}}^{\textrm{(C)}} \cdot G$;
the six-point vertex
$\mathring{\gamma}_{\bar{t}}^{\textrm{(C)}}$
needed for this
can be constructed \textit{iteratively}.
To elaborate this point, recall that the four-point vertex $\dot{\gamma}_{\bar{t}}^{\textrm{(C)}}$ constitutes a certain part of the vertex flow \eqref{eq:gamma_flow}, which can be computed in a iterative one-loop fashion. To generate the six-point vertex $\mathring{\gamma}_{\bar{t}}^{\textrm{(C)}}$, one simply has to remove the differentiated line, $\dot{G}$, in this construction: One starts from a six-point vertex obtained by removing the differentiated line in the \textit{one-loop part} of Eq.~\eqref{eq:gamma_flow}. Let us call the resulting object from the $p$ channel $\mathring{\gamma}_p^{(1)}$. Then, $\mathring{\gamma}_p^{(1)}$ can be inserted into the \textit{center part} of Eq.~\eqref{eq:gamma_flow} to generate a first contribution for $\mathring{\gamma}_a^{\textrm{(C)}}$. These steps are illustrated in Figs.~\ref{fig:bubble_cut}(c) to \ref{fig:bubble_cut}(e). Further contributions of $\mathring{\gamma}_r^{\textrm{(C)}}$ (for a certain channel $r$) are obtained as, e.g., $\mathring{\gamma}_r^{(1)}$ is inserted into the \textit{left}, \textit{right}, or \textit{center} parts [cf.\ Eq.~\eqref{eq:gamma_flow}] of channels $r'\neq r$ before inserting the resulting objects into $\mathring{\gamma}_r^{(\textrm{C})}$.
We remark that this scheme is directly accessible numerically by computing one-loop integral equations with six-point vertices. 
Though this will be computationally costly, it is conceptually not more complicated than computing the four-point mfRG flow.
In fact, it is not surprising that one has to deal with six-point objects to go beyond the initial parquet vertex, since the PA exhausts (by construction) all diagrams that can be obtained in an iterative one-loop computation involving only four-point objects.

\subsection{Schwinger-Dyson equation with reversed order}
Next, we show that the self-energy in the PA can equivalently be obtained from the SDE with either ordering of the involved vertices, i.e.,
\begin{equation}
\Sigma^{\textrm{PA}} = \Sigma_{\textrm{SD}}(\Gamma_0,\Gamma,G) = \Sigma_{\textrm{SD}}(\Gamma,\Gamma_0,G)
\ED
\label{eq:sigma_sd_sym}
\end{equation}
In Section \ref{sec:sde}, we have used the expression $\Sigma_{\textrm{SD}}(\Gamma_0,\Gamma,G)$ to derive the self-energy flow $\eqref{eq:sigma_dot_g_dot}$, which finally yielded Eq.~\eqref{eq:dsigma_PA} for the functional derivative in the PA. If we use the SDE in the ``reversed'' order, we can actually follow these steps in close analogy to find the same relation for the functional derivative.
First, starting from $\Sigma = \Sigma_{\textrm{SD}}(\Gamma,\Gamma_0,G)$, we find a 
replication of Eq.~\eqref{eq:sigma_dot_1} with reversed order:
\begin{align}
\dot{\Sigma}_1 
& =
-I_t \cdot \dot{G} +
( \Gamma \circ \dot{\Pi} \circ \gamma_a + \Gamma \circ \dot{\Pi}_p \circ \gamma_p ) \cdot G
\ED
\end{align}
Concerning the simplifications of $\dot{\Sigma}_2$, we start from
$(\dot{\Gamma} \circ \Pi_p \circ R) \cdot G$ to get [instead of Eq.~\eqref{eq:sigma_dot_2_sum}]
\begin{equation}
\dot{\Sigma}_2 = -\sum_{r=a,p}
\big(
\dot{\gamma}_r \circ \Pi_r \circ I_r
-
\dot{I}_r \circ \Pi_r \circ \gamma_r
\big)
\cdot G 
\ED
\end{equation}
Then, we use the BSE with ``reversed'' order, $\gamma_r = \Gamma \circ \Pi_r \circ I_r$ [cf.\ Eq.~\eqref{eq:rbse}], to find the appropriate version of Eq.~\eqref{eq:sigma_dot_2_final},
\begin{equation}
\dot{\Sigma}_2 = -\sum_{r=a,p} \big( \dot{\gamma}_r^{(\textrm{C})} + \Gamma \circ \dot{\Pi}_r \circ \gamma_r \big) \cdot G
\ED
\end{equation}
The final manipulations can be made in complete analogy to obtain
\begin{align}
\dot{\Sigma}^{\textrm{PA}} = 
\dot{\Sigma}_1 + \dot{\Sigma}_2
& =
-I_t \cdot \dot{G}
-\dot{\gamma}_{\bar{t}}^{\textrm{(C)}} \cdot G
\quad
\Rightarrow
\quad
\frac{\delta \Sigma^{\textrm{PA}}}{\delta G}
=
- I_t 
- \mathring{\gamma}_{\bar{t}}^{\textrm{(C)}} \cdot G
\EC
\end{align}
i.e., the identical differential equation \eqref{eq:dsigma_PA}.
Since, for the specific propagator $G=0$, one has $\Sigma_{\textrm{SD}}(\Gamma_0,\Gamma,0) = 0 =  \Sigma_{\textrm{SD}}(\Gamma,\Gamma_0,0)$, it follows that the self-energy in the PA can indeed be obtained from any of the two versions of the SDE.
The strategy of generating, first, a self-energy via the SDE and, then, obtaining a vertex by functional differentiation has been famously put forward by Baym and Kadanoff \cite{Baym1961}. 
They showed that, if the self-energy can equivalently be constructed via the SDE with either order of the vertices, then, the one-particle propagator is conserving. 
Thus, using this argument together with Eq.~\eqref{eq:sigma_sd_sym}, one finds that the PA fulfills one-particle conservation laws. 
Baym and Kadanoff further showed that, if the vertices are subsequently constructed from $I_t' = - \delta \Sigma / \delta G$ and $\Gamma' = I_t' + I_t' \circ \Pi_t \circ \Gamma'$, two-particle conservation laws are fulfilled as well. As is well known, the PA does not fulfill two-particle conservation laws. In fact, Eq.~\eqref{eq:dsigma_PA} shows how the parquet vertex $I_t$ needs to be modified to be conserving; in other words, the correction term $\mathring{\gamma}_{\bar{t}}^{\textrm{(C)}} \cdot G$ allows one to quantify to what degree the vertex $I_t$ in the PA violates conservation laws.
Furthermore, Eq.~\eqref{eq:dsigma_PA} provides a construction how to generate a fully conserving solution originating from the parquet self-energy. After both the vertex $I_t$ and the self-energy $\Sigma^{\textrm{PA}}$ in the PA have been obtained, one computes $\mathring{\gamma}_{\bar{t}}^{\textrm{(C)}} \cdot G$ and adds this to $I_t$ to get a conserving vertex $I_t'$. Note that the original parquet self-energy need not be modified. Similarly as one computes $\Gamma' = I_t' + I_t' \circ \Pi_t \circ \Gamma'$ with the original $\Pi_t$ (containing $\Sigma^{\textrm{PA}}$), physical quantities (such as susceptibilities, conductivities, etc.) are computed using $I_t'$ (or $\Gamma'$) together with $\Sigma^{\textrm{PA}}$.
The resulting solution fulfills one- and two-particle conservation laws, but, clearly, it does not fulfill the SDE anymore. This is not surprising since, as shown in Appendix \ref{sec:sigmaproof}, a solution that fulfills both the SDE and the functional derivative must be the exact solution. The preferential choice between $\Gamma$ and $\Gamma'$ will surely depend on the physical application.

\begin{figure}[t]
\includegraphics[width=0.98\textwidth]{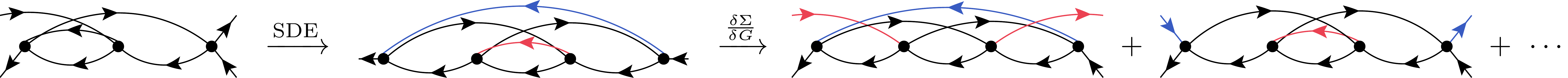}
\caption{%
Illustration for the relation between (skeleton) diagrams of the vertex and the self-energy at fourth order in the interaction: Inserting the first (parquet) vertex diagram into the Schwinger-Dyson equation, we generate the second diagram as part of $\Sigma^{\textrm{PA}}$. Upon taking the functional derivative w.r.t.\ to the full propagator, this self-energy diagram relates to multiple diagrams of the two-particle-irreducible vertex $I_t$. Among those, the third diagram, obtained by cutting the (light) red line, is an envelope diagram and not part of $I_t$ in the PA. However, the fourth diagram, obtained by cutting the blue line, belongs to it. Note that we ignore signs and prefactors in these diagrams.%
}
\label{fig:examples}
\end{figure}

We remark that there have also been suggestions of how to keep the vertex $I_t$ in the PA but modify the self-energy, $\Sigma^{\textrm{PA}}$, to obtain a thermodynamically consistent description \cite{Janis2017}. While these ideas might be useful in practical situations, it is, however, not possible to construct a combination of the skeleton two-particle-irreducible vertex $I_t[G]$ in the PA together with \textit{any} skeleton self-energy $\tilde{\Sigma}[G]$, such that the functional derivative $I_t = - \delta \tilde{\Sigma} / \delta G$ is fulfilled. The reason is that the functional derivative generates from any diagram of $\tilde{\Sigma}$ a multitude of diagrams for $I_t$---the same self-energy diagram related to missing diagrams of $I_t$ in the PA also relates to diagrams that are contained in $I_t$ (cf.\ Fig.~\ref{fig:examples}). Therefore, the functional derivative cannot be fulfilled by starting from the PA and simply removing diagrams from the self-energy.

\section{Response functions}
\label{sec:response}
Finally, we use our results from Section \ref{sec:vertex} to derive dependent, mfRG flow equations for response functions.
In fact, 
the (fermionic) four-point vertex, $\Gamma$, and the self-energy, $\Sigma$,
give us full control over correlation functions up to the four-point level,
and thus they suffice 
to compute response functions such as three-point vertices, $\Gamma^{(3)}$, and susceptibilities, $\chi$.
If $\Gamma$ and $\Sigma$ are obtained by an RG flow, the response functions can be deduced from the scale-dependent $\Gamma^{\Lambda}$, $\Sigma^{\Lambda}$ at any stage during the flow.
Alternatively, the response functions $\Gamma^{(3),\Lambda}$ and $\chi^{\Lambda}$ are often deduced from their own RG flows \cite{Metzner2012}. 
In this case, the flow equations provided by the standard fRG hierarchy again require knowledge
about unknown, higher-point vertices (namely a five-point vertex for the flow of $\Gamma^{(3)}$ and a boson-fermion four-point vertex for $\chi$) \cite{Kopietz2010}. 
In particular, the inevitable truncation in the fRG hierarchy leads to ambiguities in the computation
of the response function \cite{Kugler2018b,Tagliavini2018}.
These ambiguities have been recently resolved by a diagrammatic derivation of the mfRG flow equations for the response functions \cite{Tagliavini2018}. Here, we provide algebraic derivations of these flow equations. We find that one can circumvent the influence of unknown, higher-point vertices by using exact flow equations for the response functions, which follow from the standard relations between the response functions and the (known) fermionic four-point vertex and self-energy. 
\subsection{Three-point vertex}
\begin{figure}[t]
\includegraphics[width=.6\textwidth]{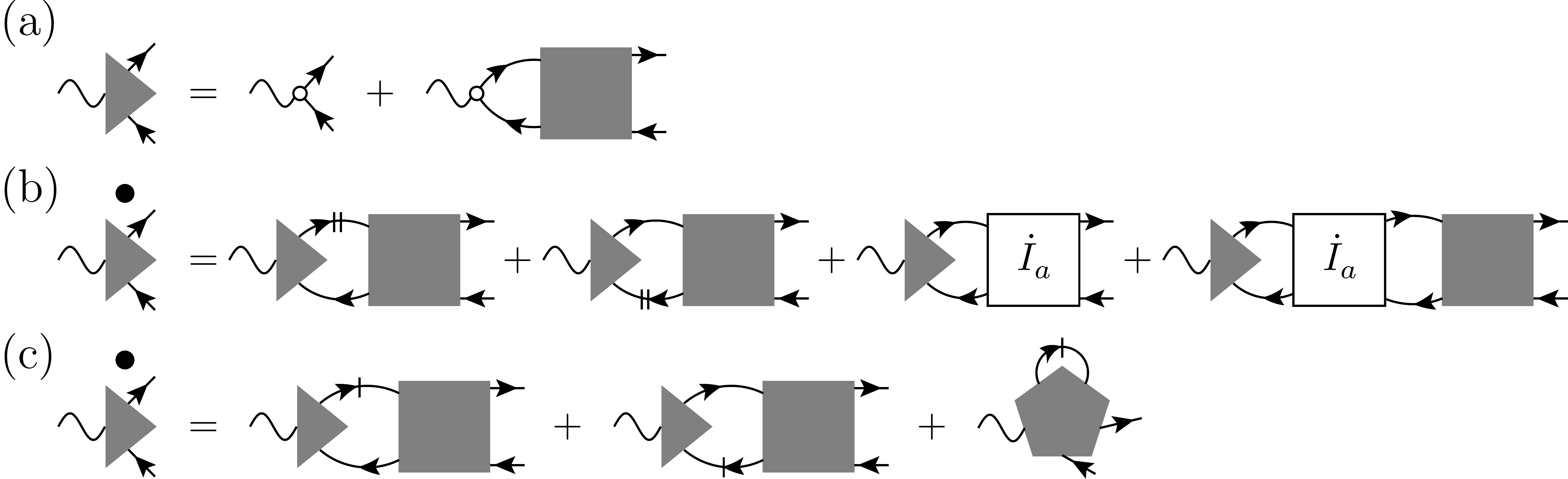}
\caption{%
Illustration of three exact equations for the three-point vertex in the $a$ channel:
(a)
Schwinger-Dyson equation between three- and four-point vertex; the white dot denotes the bare three-point vertex;
(b)
mfRG flow equation containing differentiated vertices from the complementary channel, $\dot{I}_a = \dot{\gamma}_{\bar{a}}$; and
(c)
standard fRG flow equation containing an unknown five-point vertex.%
}
\label{fig:gamma3}
\end{figure}
The Schwinger-Dyson equation relating the (full) three-point vertex 
to the bare three-point vertex (often taken to be unity) and the four-point vertex \cite{Kopietz2010}
is given by (cf.\ Fig.~\ref{fig:gamma3})
\begin{equation}
\Gamma^{(3)}_r = 
\Gamma^{(3)}_{r,0} +
\Gamma^{(3)}_{r,0} \circ \Pi_r \circ \Gamma
\label{eq:gamma3_sde}
\ED
\end{equation}
Employing the scale dependence described in the previous sections,
we can differentiate Eq.~\eqref{eq:gamma3_sde} to get
\begin{align}
\dot{\Gamma}^{(3)}_r
& =
\Gamma^{(3)}_{r,0} \circ \dot{\Pi}_r \circ \Gamma
+ \Gamma^{(3)}_{r,0} \circ \Pi_r \circ 
\dot{\Gamma}
=
\Gamma^{(3)}_{r,0} \circ \dot{\Pi}_r \circ \Gamma
+ \Gamma^{(3)}_{r,0} \circ \Pi_r \circ 
(\dot{I}_r + \dot{\gamma}_r)
\ED
\end{align}
We insert the mfRG vertex flow \eqref{eq:gamma_flow}, combine several terms according to Eq.~\eqref{eq:gamma3_sde}, and obtain
\begin{align}
\dot{\Gamma}^{(3)}_r
& =
\Gamma^{(3)}_{r,0} \circ \dot{\Pi}_r \circ \Gamma
+ \Gamma^{(3)}_{r,0} \circ \Pi_r \circ \dot{I}_r
+ \Gamma^{(3)}_{r,0} \circ \Pi_r \circ 
\big(
\Gamma \circ \dot{\Pi}_r \circ \Gamma
+ 
\Gamma \circ \Pi_r \circ \dot{I}_r
+ 
\dot{I}_r \circ \Pi_r \circ \Gamma
+
\Gamma \circ \Pi_r \circ \dot{I}_r \circ \Pi_r \circ \Gamma
\big)
\nonumber \\
& =
\Gamma^{(3)}_{r} \circ \dot{\Pi}_r \circ \Gamma
+ \Gamma^{(3)}_{r} \circ \Pi_r \circ 
\big( \dot{I}_r + \dot{I}_r \circ \Pi_r \circ \Gamma \big)
\ED
\label{eq:gammadot3}
\end{align}
The first term occurs similarly in the fRG flow equation (with the typical replacement $\dot{G} \leftrightarrow S$). However, the remaining part of our flow equation successfully replaces the contributions from the unknown five-point vertex in the fRG flow. 
\subsection{Susceptibility}
\begin{figure}[t]
\includegraphics[width=.475\textwidth]{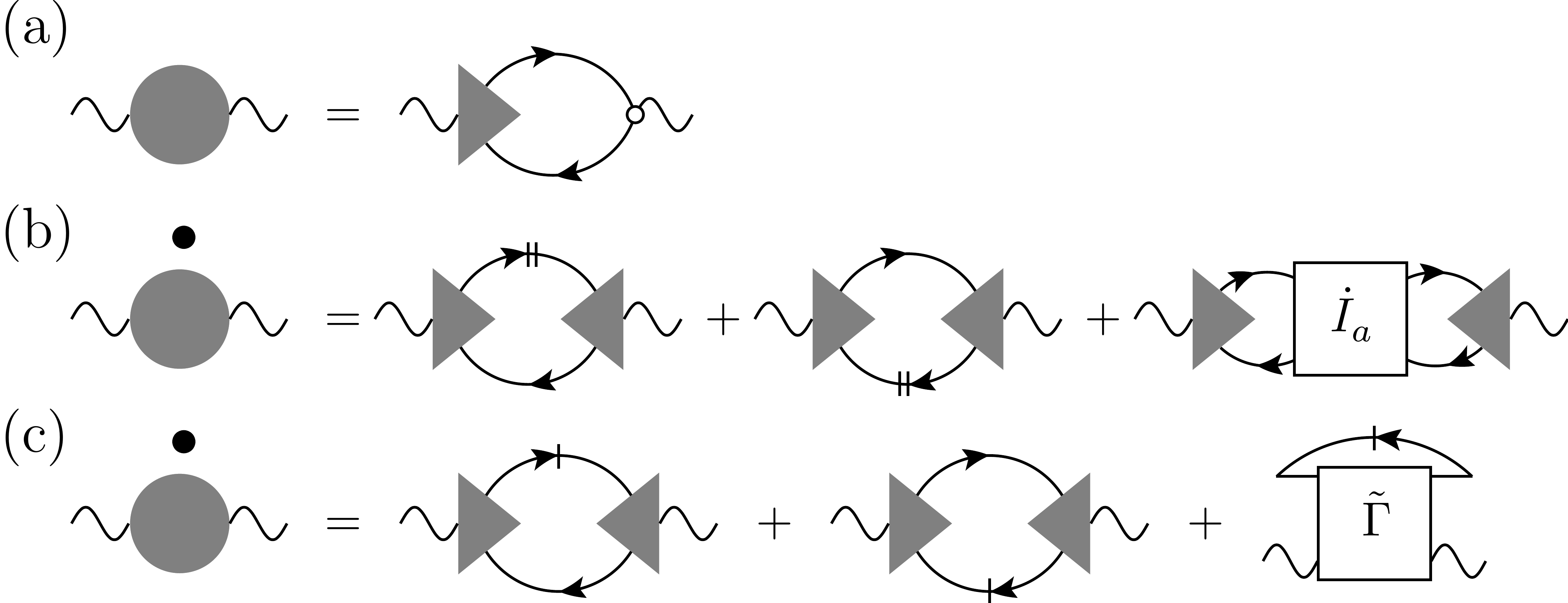}
\caption{%
Illustration of three exact equations for the susceptibility in the $a$ channel:
(a)
Schwinger-Dyson equation relating the susceptibility to the full and bare three-point vertex;
(b)
mfRG flow equation containing differentiated vertices from the complementary channel, $\dot{I}_a = \dot{\gamma}_{\bar{a}}$; and
(c)
standard fRG flow equation containing an unknown fermion-boson four-point vertex $\tilde{\Gamma}$.%
}
\label{fig:chi}
\end{figure}
The susceptibility is fully determined by the
three-point vertex or [via Eq.~\eqref{eq:gamma3_sde}]
the four-point vertex \cite{Kopietz2010}, according to (cf.\ Fig.~\ref{fig:chi})
\begin{align}
\chi_r & 
=
\Gamma^{(3)}_{r} \circ \Pi_r \circ \Gamma^{(3)\dag}_{r,0}
=
\Gamma^{(3)}_{r,0} \circ \Pi_r \circ \Gamma^{(3)\dag}_{r,0}
+
\Gamma^{(3)}_{r,0} \circ \Pi_r \circ \Gamma \circ \Pi_r \circ \Gamma^{(3)\dag}_{r,0}
\ED
\end{align}
We can differentiate either relation; choosing the first one, we insert the mfRG flow \eqref{eq:gammadot3} of $\Gamma^{(3)}$ to find the mfRG flow of the susceptibility:
\begin{align}
\dot{\chi}_r
& 
=
\Gamma^{(3)}_{r} \circ \dot{\Pi}_r \circ \Gamma^{(3)\dag}_{r,0}
+ \dot{\Gamma}^{(3)}_{r} \circ \Pi_r \circ \Gamma^{(3)\dag}_{r,0}
\nonumber \\
& =
\Gamma^{(3)}_{r} \circ \dot{\Pi}_r \circ \Gamma^{(3)\dag}_{r,0}
+
\big(
\Gamma^{(3)}_{r} \circ \dot{\Pi}_r \circ \Gamma \circ 
+ \Gamma^{(3)}_{r} \circ \Pi_r \circ \dot{I}_r
+ \Gamma^{(3)}_{r} \circ \Pi_r \circ \dot{I}_r \circ \Pi_r \circ \Gamma
\big)
\circ \Pi_r \circ \Gamma^{(3)\dag}_{r,0}
\nonumber \\
& =
\Gamma^{(3)}_{r} \circ \dot{\Pi}_r \circ \Gamma^{(3)\dag}_{r}
+ \Gamma^{(3)}_{r} \circ \Pi_r \circ \dot{I}_r \circ \Pi_r \circ \Gamma^{(3)\dag}_{r}
\ED
\label{eq:dotchi}
\end{align}
Again, the first term occurs similarly in the fRG flow equation (with $\dot{G} \leftrightarrow S$), and the remaining terms in our flow equation replace the contributions from the unknown boson-fermion four-point vertex in the fRG flow. 
Let us briefly summarize:
The response functions $\Gamma^{(3)}$, $\chi$ can be deduced from the four-point vertex, $\Gamma$, and the self-energy, $\Sigma$, at any point of the RG fow. As $\Gamma$ and $\Sigma$ evolve with $\Lambda$, so do $\Gamma^{(3)}$ and $\chi$. With the above derivation, we have cast this evolution into exact, mfRG flow equations for the response function, each containing the vertex flow from the complementary channel ($\dot{I}_r = \dot{\gamma}_{\bar{r}}$).
The two-particle-reducible vertices still obey the mfRG flow \eqref{eq:gamma_flow};
approximations come from the chosen expression for the totally irreducible vertex, $R$, which affects the initial conditions but is itself not part of the flow.
\section{Conclusion}
\label{sec:conclusion}
We have used the well-known self-consistent relations of the parquet formalism to derive exact flow equations for various vertex and correlation functions. Compared to the standard fRG framework, these multiloop fRG (mfRG) flow equations can be advantageous as they circumvent the reliance on higher-point vertices. In fact, our calculations include concise, algebraic derivations of the mfRG flow equations that have previously been derived diagrammatically \cite{Kugler2018,Kugler2018a,Tagliavini2018} 
and have already been used \cite{Kugler2018,Tagliavini2018} to improve the approximations of the truncated fRG 
flow (for results of two-loop fRG, see Refs.~\onlinecite{Eberlein2014,Wentzell2016,Rueck2018}).
The analysis presented in this paper puts the mfRG approach on a general basis. The algebraic derivations open the route to RG flows beyond the diagrams of the parquet approximation (PA). Since the totally irreducible vertex, $R$, is precisely the part of the vertex that cannot be efficiently included in the flow, the focus can now shift to systematic ways of computing $R$. 
If one chooses a scale dependence in the propagators that starts from $G_0^{\Lambda_i}=0$, all reducible contributions built on $R$ will be fully included by the mfRG flow. Other starting points for the flow are a possible as well. In particular, if one uses as initial, bare propagator the (self-consistently determined) one from dynamical mean-field theory, $G_0^{\Lambda_i}=G_0^{\textrm{DMFT}}$, the nonlocal correlations not contained in DMFT will be added by a flow that starts from the self-energy $\Sigma^{\textrm{DMFT}}$ and the full vertex $\Gamma^{\textrm{DMFT}}$ \cite{Taranto2014}, thus circumventing potential divergences of $R^{\textrm{DMFT}}$.
Similarly, if the system in question is related to another, solvable reference system \cite{Wentzell2016} by variation of one-particle parameters, mfRG can be used to tune between these systems via $G_0^{\Lambda}$, with the guarantee that the self-consistent parquet equations are fulfilled throughout the flow. 
As examples, let us mention Fermi polarons \cite{Mitra1987,Schmidt2018}, where one can tune the chemical potential of the majority species, and nonequilibrium transport (see below), where one can gradually increase the bias voltage.
Our computations also provide a basis for setting up mfRG flows for more complicated theories, including, for instance, further bosonic degrees of freedom. Generally, we believe that the insights presented in this paper will be useful for further development of quantum-field-theoretical RG techniques.

Additionally, we have demonstrated an intimate relation between the functional derivative of the self-energy (inducing a conserving solution) and the (standard) fRG self-energy flow: The flow equation directly follows from the functional derivative for the case that the propagator is varied through a scale parameter. However, a solution of the fRG flow is consistent with the functional derivative only if the flow is independent of the specific scale dependence, i.e., only if $\Gamma \cdot S$ constitutes a total derivative of diagrams. A simple example for which this is indeed the case is given by a truncated fRG flow with a (particle-hole) ladder vertex that reproduces self-consistent Hartree-Fock. Building on this, it would be worthwhile to devise other approximate flows that comply with the functional derivative but go beyond Hartree-Fock, thereby including an interplay between different two-particle channels.

Lastly, we have used our approach to address important general questions of (traditional) parquet theory. Using an argument of Baym and Kadanoff \cite{Baym1961}, we have demonstrated that the PA fulfills one-particle conservation laws. Furthermore, we have shown how to \textit{construct} a two-particle-irreducible vertex equivalent to taking the functional derivative of the parquet self-energy. With this, one can quantify to what extent the PA violates two-particle conservation laws, and one can modify the PA to obtain a fully conserving approximation. It would be interesting to apply this modified parquet approach in situations where conservation properties are crucial, such as studies of transport phenomena.

The generality of our formalism opens up a vast field of applications.
Multiloop fRG flows have already yielded impressive results for the prototypical 2D Hubbard model \cite{Tagliavini2018} (see Ref.~\onlinecite{Eberlein2014} for results using two-loop fRG) and promise a better understanding of strongly correlated electron systems \cite{Kugler2018a,Metzner2012,Rohringer2017}.
In the study of quantum magnetism, the pseudo-fermion fRG approach \cite{Reuther2010} has become a competing method, and first calculations with two-loop corrections \citep{Rueck2018} suggest that a full multiloop treatment would yield further improvements.
Moreover, mfRG can be directly applied to a variety of interesting physical problems where the most relevant properties are expected to emerge within the PA, such as various forms of mobile impurity problems \cite{Schmidt2018,Schmidt2011} or one-dimensional fermion systems \cite{Meden2008} beyond the Luttinger liquid paradigm \cite{Imambekov2012}.
In the field of transport phenomena in disordered systems, our mfRG approach could provide unprecedented insight into many-body localization in large systems \cite{Abanin2017,Pietracaprina2018} or interaction effects on the Anderson localization transition \cite{Amini2014}.
Finally, we remark that mfRG flows can also be naturally set up within the Keldysh formalism \cite{Jakobs2010,Jakobs2010a} to provide real-frequency information, both in and out of equilibrium.
 
\begin{acknowledgments}
We thank S.\ Andergassen, P.\ Chalupa, A.\ Tagliavini, and A.\ Toschi
for useful discussions at TU Wien. We thank O.\ Yevtushenko as well as A.\ Toschi for a careful reading of the manuscript and A.\ A.\ Katanin for a helpful correspondence.
We acknowledge support by the Cluster of Excellence
Nanosystems Initiative Munich;
F.B.K.\ acknowledges funding from
the research school IMPRS-QST.
\end{acknowledgments}

\appendix
\section{Matrix notation of bubbles and loops}
\label{sec:matrixnotation}
\begin{figure}[t]
\includegraphics[width=.98\textwidth]{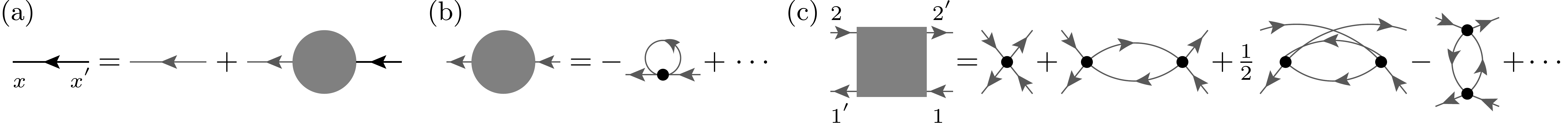}
\caption{%
(a)
Dyson's equation relating the full propagator $G_{x,x\p}$ (black, thick line)
to the bare propagator $G_0$ (gray, thin line)
and the self-energy $\Sigma$ (circle).
(b)
First-order diagram for the self-energy
using the (antisymmetrized) bare vertex $\Gamma_0$ (solid dot).
(c)
Diagrammatic expansion of the four-point vertex
$\Gamma$ (square)
up to second order in the interaction.
The positions of the external (amputated) legs
refer to the arguments of 
$\Gamma_{x_1\p, x_2\p; x_1\ppp, x_2\ppp}$.
Diagrams from left to right belong to $R$, $\gamma_a$, $\gamma_p$, and $\gamma_t$, respectively; diagrams for $I_r$ follow from the relation $I_r = \Gamma - \gamma_r$.%
}
\label{fig:indices}
\end{figure}

In this section, we define our notation for the contraction of various vertex functions.
It is common to view the contraction of one-particle quantities as matrix multiplications, such that, e.g., the Dyson equation between propagator, $G_{x, x\p} = - \langle c_{x} \bar{c}_{x\p} \rangle$, and the self-energy, $\Sigma_{x\p,x}$, [cf.\ Figs.~\ref{fig:indices}(a) and \ref{fig:indices}(b)] reads
\begin{equation}
G = G_0 + G_0 \cdot \Sigma \cdot G
\EC \quad
( A \cdot B )_{x, x\p} = \sum_{y} A_{x, y} B_{y, x\p}
\ED
\end{equation}
For the contraction of two four-point vertices, we have three inequivalent possibilities corresponding to the three two-particle channels $r=a,p,t$ [standing for antiparallel, parallel, transverse, respectively; cf.\ also Fig.~\ref{fig:indices}(c)].
In Ref.~\onlinecite{Kugler2018a}, the different combinations have been labeled as ``bubble functions'' $B_r(\Gamma,\Gamma')$. Here, we repeat the corresponding equations and show that they can be conveniently written as matrix multiplications. For this, we will use auxiliary objects that depend on channel-dependent tuples of quantum numbers (e.g., $\Gamma_{x_1\p, x_2\p; x_1\pp, x_2\pp} = \tilde{\Gamma}_{a;(x_1\p, x_2\pp), (x_2\p, x_1\pp)} $) and define a contraction $\circ$ that always comes together with a two-particle propagator $\Pi_r$ of a certain channel (consisting of two one-particle propagators $G$):
\begin{subequations}
\begin{align}
B_a(\Gamma, \Gamma')_{x_1\p, x_2\p; x_1\pp, x_2\pp} 
& =
\sum_{y_1\p, y_1\pp, y_2\p, y_2\pp} 
\Gamma_{x_1\p, y_2\p; y_1\pp, x_2\pp}
G_{y_1\pp, y_1\p} G_{y_2\pp, y_2\p} 
\Gamma'_{y_1\p, x_2\p; x_1\pp, y_2\pp}
\nonumber \\
& =
\sum_{y_1\p, y_1\pp, y_2\p, y_2\pp} 
\tilde{\Gamma}_{a;(x_1\p,x_2\pp),(y_2\p,y_1\pp)}
\tilde{\Pi}_{a;(y_2\p,y_1\pp),(y_1\p,y_2\pp)} 
\tilde{\Gamma}'_{a;(y_1\p,y_2\pp),(x_2\p,x_1\pp)}
\equiv
( \Gamma \circ \Pi_a \circ \Gamma')_{x_1\p, x_2\p; x_1\pp, x_2\pp} 
\EC
\label{eq:abubble} \\
B_p(\Gamma, \Gamma')_{x_1\p, x_2\p; x_1\pp, x_2\pp}
& = \tfrac{1}{2}
\sum_{y_1\p, y_1\pp, y_2\p, y_2\pp} 
\Gamma_{x_1\p, x_2\p; y_1, y_2\pp}
G_{y_1\pp, y_1\p} G_{y_2\pp, y_2\p} 
\Gamma'_{y_1\p, y_2\p; x_1\pp, x_2\pp}
\nonumber \\
& = 
\sum_{y_1\p, y_1\pp, y_2\p, y_2\pp} 
\tilde{\Gamma}_{p;(x_1\p,x_2\p),(y_1\pp,y_2\pp)}
\tilde{\Pi}_{p;(y_1\pp,y_2\pp),(y_1\p,y_2\p)}
\tilde{\Gamma}'_{p;(y_1\p,y_2\p),(x_1\pp,x_2\pp)}
\equiv
( \Gamma \circ \Pi_p \circ \Gamma')_{x_1\p, x_2\p; x_1\pp, x_2\pp} 
\EC
\label{eq:pbubble} \\
B_t(\Gamma, \Gamma')_{x_1\p, x_2\p; x_1\pp, x_2\pp} 
& = -
\sum_{y_1\p, y_1\pp, y_2\p, y_2\pp} 
\Gamma_{y_1\p, x_2\p; y_1\pp, x_2\pp}
G_{y_2\pp, y_1\p} G_{y_1\pp, y_2\p}
\Gamma'_{x_1\p, y_2\p; x_1\pp, y_2\pp}
\nonumber \\
& = 
\sum_{(y_1\p, y_1\pp),(y_2\p, y_2\pp)}
\tilde{\Gamma}_{t;(x_2\p,x_2\pp),(y_1\p,y_1\pp)}
\tilde{\Pi}_{t;(y_1\p,y_1\pp),(y_2\p,y_2\pp)}
\tilde{\Gamma}'_{t;(y_2\p,y_2\pp),(x_1\p,x_1\pp)}
\equiv
( \Gamma \circ \Pi_t \circ \Gamma' )_{x_1\p, x_2\p; x_1\pp, x_2\pp} 
\ED
\label{eq:tbubble}
\end{align}
\label{eq:bubbles}
\end{subequations}
Note that a factor of 1/2 has been absorbed into $\Pi_p$ and a minus sign into $\Pi_t$.
From Eqs.~\eqref{eq:action} and \eqref{eq:four-point_correlator}, it is clear that $\Gamma_0$ and $\Gamma$ are antisymmetric in their indices. Using the bubble functions \eqref{eq:bubbles} together with the parquet equations \eqref{eq:parquet}, one finds the further crossing symmetries stated in Eq.~\eqref{eq:crossing}, which use the symbol
\begin{equation}
\hat{\Gamma}_{x_1\p, x_2\p; x_1\pp, x_2\pp} 
=
\Gamma_{x_1\p, x_2\p; x_2\pp, x_1\pp} 
=
\Gamma_{x_2\p, x_1\p; x_1\pp, x_2\pp} 
\ED
\label{eq:exchange_legs}
\end{equation}
If we combine two fermionic indices into one bosonic index, the above equations directly translate to three-point vertices. For instance, one could combine the two external legs of the first vertex in the $a$ bubble according to some function $f$ and interpret
\begin{equation}
\Gamma^{(3)}_{a;z,x_2\p,x_1\pp} = 
\sum_{x_1\p,x_2\pp} f_{z,x_1\p,x_2\pp} \Gamma_{x_1\p, y_2\p; y_1\pp, x_2\pp}
\quad \Rightarrow \quad
( \Gamma^{(3)}_a \circ \Pi_a \circ \Gamma')_{z,x_2\p,x_1\pp}
=
\sum_{x_1\p,x_2\pp} f_{z,x_1\p,x_2\pp}
(\Gamma \circ \Pi_a \circ \Gamma')_{x_1\p, x_2\p; x_1\pp, x_2\pp}
\ED
\end{equation}
Furthermore, one can contract a four-point vertex with a one-particle propagator to obtain another one-particle object. We define the symbol $\cdot$ between vertex and propagator to be such a contraction applied to the ``upper'' external legs of the vertex [i.e., legs $2$ and $2'$ in Fig.~\ref{fig:indices}(c)]. In Ref.~\onlinecite{Kugler2018a}, this has been dubbed a ``self-energy loop'', $L$, defined as
\begin{align}
- L(\Gamma, G)_{x\p,x\pp}
=
\sum_{y\p, y} \Gamma_{x\p, y\p; x, y} G_{y, y\p}
=
\sum_{(y\p, y)} \tilde{\Gamma}_{t;(x\p,x),(y\p,y)} \tilde{G}_{(y\p,y)}
\equiv
( \Gamma \cdot G )_{x\p,x\pp}
\ED
\label{eq:self-energy_loop}
\end{align}
If the contracting line is a composite object of the type $G \cdot \Sigma \cdot G$,
we can view the $G$ lines as a $t$ bubble attached to the vertex, according to
\begin{align}
\big( \Gamma \cdot ( G\cdot \Sigma \cdot G) \big)_{x\p, x}
& =
\sum_{y\p, y, z\p, z} \Gamma_{x\p, y\p; x, y} G_{y, z\p} \Sigma_{z\p,z} G_{z, y\p}
\nonumber \\
& = -
\sum_{(y\p, y), (z\p, z)} \tilde{\Gamma}_{t;(x\p,x),(y\p,y)} \tilde{\Pi}_{t;(y\p,y),(z\p,z)} 
\tilde{\Sigma}_{(z\p,z)}
\equiv -
( \Gamma \circ \Pi_t \cdot \Sigma )_{x\p, x}
\ED
\label{eq:loopbubble}
\end{align}
The Schwinger-Dyson equation for the self-energy contains a contraction of three propagators. Using the bubble functions defined above, this can equivalently be written with $\Pi_p$ and $\Pi_a$:
\begin{align}
- \Sigma_{x\p, x}
& = 
\sum_{y\p, y} \Gamma_{0;x\p, y\p; x, y} G_{y, y\p}
+
\tfrac{1}{2} \sum_{y\p, y, z\p, z, w\p, w}
\Gamma^0_{x\p, z\p; y, w}
G_{y, y\p} G_{z, z\p} G_{w, w\p}
\Gamma_{y\p, w\p; x, z}
\nonumber \\
&
= \big( ( \Gamma_0 + \Gamma_0 \circ \Pi_p \circ \Gamma ) \cdot G \big)_{x\p, x}
= \big( ( \Gamma_0 + \tfrac{1}{2} \Gamma_0 \circ \Pi_a \circ \Gamma) \cdot G \big)_{x\p, x}
\ED
\label{eq:SchwingerDyson}
\end{align}
The functional derivative between self-energy and two-particle-irreducible vertex (in the $t$ or $a$ channel) is given by
\begin{equation}
\frac{\delta \Sigma_{x\p,x}}{\delta G_{y,y\p}} =
-I_{t; x\p,y\p;x,y} = I_{a; x\p,y\p;y,x}
\ED
\label{eq:FunctionalDerivative}
\end{equation}
Note that in order to obtain the two-particle-irreducible vertex in the $p$ channel from functional differentiation, $I_{p; x\p,y\p;x,y} = \delta \Sigma_{x\p,y\p}/\delta G_{x,y}$,
one has to allow for variations around the physical solution which break charge conservation.
\section{Schwinger-Dyson equation and functional derivative}
\label{sec:sigmaproof}

We consider the Schwinger-Dyson equation (SDE) for the self-energy as well as the functional derivative [cf.\ Eq.~\eqref{eq:FunctionalDerivative}] between self-energy and vertex,
\begin{subequations}
\begin{align}
\Sigma & = - \Gamma_0 \cdot G -  (\Gamma_0 \circ \Pi_p \circ \Gamma) \cdot G
\EC 
\label{eq:app:sde}
\\
I_t & = - \frac{\delta \Sigma}{\delta G}
\EC \quad
\Gamma = I_t + I_t \circ \Pi_t \circ \Gamma
\EC
\label{eq:app:fd}
\end{align}
\label{eq:app:sdeandfd}%
\end{subequations}
and show that a solution for $\Sigma$ and $\Gamma$ that fulfills both Eqs.~\eqref{eq:app:sde} and \eqref{eq:app:fd} must necessarily be the exact solution. In essence, this proof has already been given by Smith \cite{Smith1992}. However, we find it useful to present it here in our notation, which exclusively consists of properly symmetrized objects.
In fact, this proof puts on solid ground what has long been known to the community \cite{Bickers2004}: In any approximate solution to the many-body problem, one has to decide whether to comply with \textit{either} conservation laws \textit{or} crossing symmetry; achieving both amounts to finding the exact solution.

To be able to apply the functional derivative, we consider the self-energy as a functional of the full propagator, $\Sigma[G]$. This is perfectly compatible with the SDE \eqref{eq:app:sde}, which is formulated using full propagators only. Furthermore, all vertex functions depend on the given theory's bare vertex $\Gamma_0$ (which we here label $\Gamma_0=U$ for ease of notation); in particular, this holds for $\Sigma[G,U]$ and $\Gamma[G,U]$. Since $U$ is the bare vertex, we have $\Gamma[G,U]=U+\mathit{O}(G^2,U^2)$; by use of either the SDE \eqref{eq:app:sde} or the functional derivative \eqref{eq:app:fd}, it is clear that
$\Sigma[G,U] = U\cdot G+\mathit{O}(G^3,U^2)$.

Assume that we know the exact vertex up to terms of order $n\geq 2$ in both $G$ and $U$, i.e.,
$\Gamma = \Gamma^{\textrm{ex}}+\mathit{O}(G^n,U^n)$.
If we apply the SDE \eqref{eq:app:sde}, we obtain (inserting into the second term)
$\Sigma = \Sigma^{\textrm{ex}}+\mathit{O}(G^{n+3},U^{n+1})$.
Now, we apply the functional derivative \eqref{eq:app:fd} and get
$I_t = I_t^{\textrm{ex}}+\mathit{O}(G^{n+2},U^{n+1})$.
Finally, using the BSE \eqref{eq:app:fd} yields
$\Gamma = \Gamma^{\textrm{ex}}+\mathit{O}(G^{n+2},U^{n+1})$,
i.e., the exact vertex one order higher in $G^2$ and $U$ than we started with.
Since we do know the exact vertex up to terms of second order, 
$\Gamma[G,U]=U+\mathit{O}(G^2,U^2)$, 
it follows by induction that a solution which fulfills both Eq.~\eqref{eq:app:sde} and \eqref{eq:app:fd} consists of the exact functionals $\Sigma^{\textrm{ex}}[G,U]$, $\Gamma^{\textrm{ex}}[G,U]$.

We remark that this proof applies equivalently to finite-order approximations of $\Sigma$ and $\Gamma$ as well as to approximations of infinite order in $U$. As soon as an expression for $\Gamma$ contains the bare vertex $U$ \cite{Smith1992}, the combination of Eq.~\eqref{eq:app:sde} and \eqref{eq:app:fd} requires \textit{all} expansion coefficients of $\Sigma$ and $\Gamma$ to be the ones of the exact solution.

\bibliographystyle{apsrev4-1}
\bibliography{references}

\begin{thebibliography}{56}%
\makeatletter
\providecommand \@ifxundefined [1]{%
 \@ifx{#1\undefined}
}%
\providecommand \@ifnum [1]{%
 \ifnum #1\expandafter \@firstoftwo
 \else \expandafter \@secondoftwo
 \fi
}%
\providecommand \@ifx [1]{%
 \ifx #1\expandafter \@firstoftwo
 \else \expandafter \@secondoftwo
 \fi
}%
\providecommand \natexlab [1]{#1}%
\providecommand \enquote  [1]{``#1''}%
\providecommand \bibnamefont  [1]{#1}%
\providecommand \bibfnamefont [1]{#1}%
\providecommand \citenamefont [1]{#1}%
\providecommand \href@noop [0]{\@secondoftwo}%
\providecommand \href [0]{\begingroup \@sanitize@url \@href}%
\providecommand \@href[1]{\@@startlink{#1}\@@href}%
\providecommand \@@href[1]{\endgroup#1\@@endlink}%
\providecommand \@sanitize@url [0]{\catcode `\\12\catcode `\$12\catcode
  `\&12\catcode `\#12\catcode `\^12\catcode `\_12\catcode `\%12\relax}%
\providecommand \@@startlink[1]{}%
\providecommand \@@endlink[0]{}%
\providecommand \url  [0]{\begingroup\@sanitize@url \@url }%
\providecommand \@url [1]{\endgroup\@href {#1}{\urlprefix }}%
\providecommand \urlprefix  [0]{URL }%
\providecommand \Eprint [0]{\href }%
\providecommand \doibase [0]{http://dx.doi.org/}%
\providecommand \selectlanguage [0]{\@gobble}%
\providecommand \bibinfo  [0]{\@secondoftwo}%
\providecommand \bibfield  [0]{\@secondoftwo}%
\providecommand \translation [1]{[#1]}%
\providecommand \BibitemOpen [0]{}%
\providecommand \bibitemStop [0]{}%
\providecommand \bibitemNoStop [0]{.\EOS\space}%
\providecommand \EOS [0]{\spacefactor3000\relax}%
\providecommand \BibitemShut  [1]{\csname bibitem#1\endcsname}%
\let\auto@bib@innerbib\@empty
\bibitem [{\citenamefont {Hedin}(1965)}]{Hedin1965}%
  \BibitemOpen
  \bibfield  {author} {\bibinfo {author} {\bibfnamefont {L.}~\bibnamefont
  {Hedin}},\ }\href {\doibase 10.1103/PhysRev.139.A796} {\bibfield  {journal}
  {\bibinfo  {journal} {Phys. Rev.}\ }\textbf {\bibinfo {volume} {139}},\
  \bibinfo {pages} {A796} (\bibinfo {year} {1965})}\BibitemShut {NoStop}%
\bibitem [{\citenamefont {Bickers}(2004)}]{Bickers2004}%
  \BibitemOpen
  \bibfield  {author} {\bibinfo {author} {\bibfnamefont {N.}~\bibnamefont
  {Bickers}},\ }in\ \href
  {http://link.springer.com/chapter/10.1007%2F0-387-21717-7_6} {\emph {\bibinfo
  {booktitle} {Theoretical Methods for Strongly Correlated Electrons}}},\
  \bibinfo {series and number} {CRM Series in Mathematical Physics},\ \bibinfo
  {editor} {edited by\ \bibinfo {editor} {\bibfnamefont {D.}~\bibnamefont
  {S\'{e}n\'{e}chal}}, \bibinfo {editor} {\bibfnamefont {A.-M.}\ \bibnamefont
  {Tremblay}}, \ and\ \bibinfo {editor} {\bibfnamefont {C.}~\bibnamefont
  {Bourbonnais}}}\ (\bibinfo  {publisher} {Springer New York},\ \bibinfo {year}
  {2004})\ pp.\ \bibinfo {pages} {237--296}\BibitemShut {NoStop}%
\bibitem [{\citenamefont {Wilson}(1975)}]{Wilson1975}%
  \BibitemOpen
  \bibfield  {author} {\bibinfo {author} {\bibfnamefont {K.~G.}\ \bibnamefont
  {Wilson}},\ }\href {\doibase 10.1103/revmodphys.47.773} {\bibfield  {journal}
  {\bibinfo  {journal} {Rev. Mod. Phys.}\ }\textbf {\bibinfo {volume} {47}},\
  \bibinfo {pages} {773–840} (\bibinfo {year} {1975})}\BibitemShut {NoStop}%
\bibitem [{\citenamefont {Anderson}(1970)}]{Anderson1970}%
  \BibitemOpen
  \bibfield  {author} {\bibinfo {author} {\bibfnamefont {P.~W.}\ \bibnamefont
  {Anderson}},\ }\href {http://stacks.iop.org/0022-3719/3/i=12/a=008}
  {\bibfield  {journal} {\bibinfo  {journal} {J. Phys. C}\ }\textbf {\bibinfo
  {volume} {3}},\ \bibinfo {pages} {2436} (\bibinfo {year} {1970})}\BibitemShut
  {NoStop}%
\bibitem [{\citenamefont {Metzner}\ \emph {et~al.}(2012)\citenamefont
  {Metzner}, \citenamefont {Salmhofer}, \citenamefont {Honerkamp},
  \citenamefont {Meden},\ and\ \citenamefont {Sch\"onhammer}}]{Metzner2012}%
  \BibitemOpen
  \bibfield  {author} {\bibinfo {author} {\bibfnamefont {W.}~\bibnamefont
  {Metzner}}, \bibinfo {author} {\bibfnamefont {M.}~\bibnamefont {Salmhofer}},
  \bibinfo {author} {\bibfnamefont {C.}~\bibnamefont {Honerkamp}}, \bibinfo
  {author} {\bibfnamefont {V.}~\bibnamefont {Meden}}, \ and\ \bibinfo {author}
  {\bibfnamefont {K.}~\bibnamefont {Sch\"onhammer}},\ }\href {\doibase
  10.1103/RevModPhys.84.299} {\bibfield  {journal} {\bibinfo  {journal} {Rev.
  Mod. Phys.}\ }\textbf {\bibinfo {volume} {84}},\ \bibinfo {pages} {299}
  (\bibinfo {year} {2012})}\BibitemShut {NoStop}%
\bibitem [{\citenamefont {Kopietz}\ \emph {et~al.}(2010)\citenamefont
  {Kopietz}, \citenamefont {Bartosch},\ and\ \citenamefont
  {Sch\"{u}tz}}]{Kopietz2010}%
  \BibitemOpen
  \bibfield  {author} {\bibinfo {author} {\bibfnamefont {P.}~\bibnamefont
  {Kopietz}}, \bibinfo {author} {\bibfnamefont {L.}~\bibnamefont {Bartosch}}, \
  and\ \bibinfo {author} {\bibfnamefont {F.}~\bibnamefont {Sch\"{u}tz}},\
  }\href {http://www.springer.com/de/book/9783642050930} {\emph {\bibinfo
  {title} {Introduction to the Functional Renormalization Group}}},\ Lecture
  Notes in Physics\ (\bibinfo  {publisher} {Springer, Berlin},\ \bibinfo {year}
  {2010})\BibitemShut {NoStop}%
\bibitem [{\citenamefont {Wetterich}(1993)}]{Wetterich1993}%
  \BibitemOpen
  \bibfield  {author} {\bibinfo {author} {\bibfnamefont {C.}~\bibnamefont
  {Wetterich}},\ }\href {\doibase 10.1016/0370-2693(93)90726-x} {\bibfield
  {journal} {\bibinfo  {journal} {Phys. Lett. B}\ }\textbf {\bibinfo {volume}
  {301}},\ \bibinfo {pages} {90} (\bibinfo {year} {1993})}\BibitemShut
  {NoStop}%
\bibitem [{\citenamefont {Veschgini}\ and\ \citenamefont
  {Salmhofer}(2013)}]{Veschgini2013}%
  \BibitemOpen
  \bibfield  {author} {\bibinfo {author} {\bibfnamefont {K.}~\bibnamefont
  {Veschgini}}\ and\ \bibinfo {author} {\bibfnamefont {M.}~\bibnamefont
  {Salmhofer}},\ }\href {\doibase 10.1103/PhysRevB.88.155131} {\bibfield
  {journal} {\bibinfo  {journal} {Phys. Rev. B}\ }\textbf {\bibinfo {volume}
  {88}},\ \bibinfo {pages} {155131} (\bibinfo {year} {2013})}\BibitemShut
  {NoStop}%
\bibitem [{\citenamefont {Roulet}\ \emph {et~al.}(1969)\citenamefont {Roulet},
  \citenamefont {Gavoret},\ and\ \citenamefont {Nozi\`eres}}]{Roulet1969}%
  \BibitemOpen
  \bibfield  {author} {\bibinfo {author} {\bibfnamefont {B.}~\bibnamefont
  {Roulet}}, \bibinfo {author} {\bibfnamefont {J.}~\bibnamefont {Gavoret}}, \
  and\ \bibinfo {author} {\bibfnamefont {P.}~\bibnamefont {Nozi\`eres}},\
  }\href {\doibase 10.1103/PhysRev.178.1072} {\bibfield  {journal} {\bibinfo
  {journal} {Phys. Rev.}\ }\textbf {\bibinfo {volume} {178}},\ \bibinfo {pages}
  {1072} (\bibinfo {year} {1969})}\BibitemShut {NoStop}%
\bibitem [{\citenamefont {Shankar}(1994)}]{Shankar1994}%
  \BibitemOpen
  \bibfield  {author} {\bibinfo {author} {\bibfnamefont {R.}~\bibnamefont
  {Shankar}},\ }\href {\doibase 10.1103/RevModPhys.66.129} {\bibfield
  {journal} {\bibinfo  {journal} {Rev. Mod. Phys.}\ }\textbf {\bibinfo {volume}
  {66}},\ \bibinfo {pages} {129} (\bibinfo {year} {1994})}\BibitemShut
  {NoStop}%
\bibitem [{\citenamefont {Kugler}\ and\ \citenamefont {von
  Delft}(2018{\natexlab{a}})}]{Kugler2018}%
  \BibitemOpen
  \bibfield  {author} {\bibinfo {author} {\bibfnamefont {F.~B.}\ \bibnamefont
  {Kugler}}\ and\ \bibinfo {author} {\bibfnamefont {J.}~\bibnamefont {von
  Delft}},\ }\href {\doibase 10.1103/PhysRevLett.120.057403} {\bibfield
  {journal} {\bibinfo  {journal} {Phys. Rev. Lett.}\ }\textbf {\bibinfo
  {volume} {120}},\ \bibinfo {pages} {057403} (\bibinfo {year}
  {2018}{\natexlab{a}})}\BibitemShut {NoStop}%
\bibitem [{\citenamefont {Kugler}\ and\ \citenamefont {von
  Delft}(2018{\natexlab{b}})}]{Kugler2018a}%
  \BibitemOpen
  \bibfield  {author} {\bibinfo {author} {\bibfnamefont {F.~B.}\ \bibnamefont
  {Kugler}}\ and\ \bibinfo {author} {\bibfnamefont {J.}~\bibnamefont {von
  Delft}},\ }\href {\doibase 10.1103/PhysRevB.97.035162} {\bibfield  {journal}
  {\bibinfo  {journal} {Phys. Rev. B}\ }\textbf {\bibinfo {volume} {97}},\
  \bibinfo {pages} {035162} (\bibinfo {year} {2018}{\natexlab{b}})}\BibitemShut
  {NoStop}%
\bibitem [{\citenamefont {{Tagliavini}}\ \emph {et~al.}()\citenamefont
  {{Tagliavini}}, \citenamefont {{Hille}}, \citenamefont {{Kugler}},
  \citenamefont {{Andergassen}}, \citenamefont {{Toschi}},\ and\ \citenamefont
  {{Honerkamp}}}]{Tagliavini2018}%
  \BibitemOpen
  \bibfield  {author} {\bibinfo {author} {\bibfnamefont {A.}~\bibnamefont
  {{Tagliavini}}}, \bibinfo {author} {\bibfnamefont {C.}~\bibnamefont
  {{Hille}}}, \bibinfo {author} {\bibfnamefont {F.~B.}\ \bibnamefont
  {{Kugler}}}, \bibinfo {author} {\bibfnamefont {S.}~\bibnamefont
  {{Andergassen}}}, \bibinfo {author} {\bibfnamefont {A.}~\bibnamefont
  {{Toschi}}}, \ and\ \bibinfo {author} {\bibfnamefont {C.}~\bibnamefont
  {{Honerkamp}}},\ }\href {http://arxiv.org/abs/18070.2697} {\ }\Eprint
  {http://arxiv.org/abs/1807.02697} {arXiv:1807.02697} \BibitemShut {NoStop}%
\bibitem [{\citenamefont {Baym}\ and\ \citenamefont
  {Kadanoff}(1961)}]{Baym1961}%
  \BibitemOpen
  \bibfield  {author} {\bibinfo {author} {\bibfnamefont {G.}~\bibnamefont
  {Baym}}\ and\ \bibinfo {author} {\bibfnamefont {L.~P.}\ \bibnamefont
  {Kadanoff}},\ }\href {\doibase 10.1103/PhysRev.124.287} {\bibfield  {journal}
  {\bibinfo  {journal} {Phys. Rev.}\ }\textbf {\bibinfo {volume} {124}},\
  \bibinfo {pages} {287} (\bibinfo {year} {1961})}\BibitemShut {NoStop}%
\bibitem [{\citenamefont {Smith}(1992)}]{Smith1992}%
  \BibitemOpen
  \bibfield  {author} {\bibinfo {author} {\bibfnamefont {R.~A.}\ \bibnamefont
  {Smith}},\ }\href {\doibase 10.1103/PhysRevA.46.4586} {\bibfield  {journal}
  {\bibinfo  {journal} {Phys. Rev. A}\ }\textbf {\bibinfo {volume} {46}},\
  \bibinfo {pages} {4586} (\bibinfo {year} {1992})}\BibitemShut {NoStop}%
\bibitem [{\citenamefont {Yang}\ \emph {et~al.}(2009)\citenamefont {Yang},
  \citenamefont {Fotso}, \citenamefont {Liu}, \citenamefont {Maier},
  \citenamefont {Tomko}, \citenamefont {D'Azevedo}, \citenamefont {Scalettar},
  \citenamefont {Pruschke},\ and\ \citenamefont {Jarrell}}]{Yang2009}%
  \BibitemOpen
  \bibfield  {author} {\bibinfo {author} {\bibfnamefont {S.~X.}\ \bibnamefont
  {Yang}}, \bibinfo {author} {\bibfnamefont {H.}~\bibnamefont {Fotso}},
  \bibinfo {author} {\bibfnamefont {J.}~\bibnamefont {Liu}}, \bibinfo {author}
  {\bibfnamefont {T.~A.}\ \bibnamefont {Maier}}, \bibinfo {author}
  {\bibfnamefont {K.}~\bibnamefont {Tomko}}, \bibinfo {author} {\bibfnamefont
  {E.~F.}\ \bibnamefont {D'Azevedo}}, \bibinfo {author} {\bibfnamefont {R.~T.}\
  \bibnamefont {Scalettar}}, \bibinfo {author} {\bibfnamefont {T.}~\bibnamefont
  {Pruschke}}, \ and\ \bibinfo {author} {\bibfnamefont {M.}~\bibnamefont
  {Jarrell}},\ }\href {\doibase 10.1103/PhysRevE.80.046706} {\bibfield
  {journal} {\bibinfo  {journal} {Phys. Rev. E}\ }\textbf {\bibinfo {volume}
  {80}},\ \bibinfo {pages} {046706} (\bibinfo {year} {2009})}\BibitemShut
  {NoStop}%
\bibitem [{\citenamefont {Tam}\ \emph {et~al.}(2013)\citenamefont {Tam},
  \citenamefont {Fotso}, \citenamefont {Yang}, \citenamefont {Lee},
  \citenamefont {Moreno}, \citenamefont {Ramanujam},\ and\ \citenamefont
  {Jarrell}}]{Tam2013}%
  \BibitemOpen
  \bibfield  {author} {\bibinfo {author} {\bibfnamefont {K.-M.}\ \bibnamefont
  {Tam}}, \bibinfo {author} {\bibfnamefont {H.}~\bibnamefont {Fotso}}, \bibinfo
  {author} {\bibfnamefont {S.-X.}\ \bibnamefont {Yang}}, \bibinfo {author}
  {\bibfnamefont {T.-W.}\ \bibnamefont {Lee}}, \bibinfo {author} {\bibfnamefont
  {J.}~\bibnamefont {Moreno}}, \bibinfo {author} {\bibfnamefont
  {J.}~\bibnamefont {Ramanujam}}, \ and\ \bibinfo {author} {\bibfnamefont
  {M.}~\bibnamefont {Jarrell}},\ }\href {\doibase 10.1103/PhysRevE.87.013311}
  {\bibfield  {journal} {\bibinfo  {journal} {Phys. Rev. E}\ }\textbf {\bibinfo
  {volume} {87}},\ \bibinfo {pages} {013311} (\bibinfo {year}
  {2013})}\BibitemShut {NoStop}%
\bibitem [{\citenamefont {Li}\ \emph {et~al.}(2016)\citenamefont {Li},
  \citenamefont {Wentzell}, \citenamefont {Pudleiner}, \citenamefont
  {Thunstr\"om},\ and\ \citenamefont {Held}}]{Li2016}%
  \BibitemOpen
  \bibfield  {author} {\bibinfo {author} {\bibfnamefont {G.}~\bibnamefont
  {Li}}, \bibinfo {author} {\bibfnamefont {N.}~\bibnamefont {Wentzell}},
  \bibinfo {author} {\bibfnamefont {P.}~\bibnamefont {Pudleiner}}, \bibinfo
  {author} {\bibfnamefont {P.}~\bibnamefont {Thunstr\"om}}, \ and\ \bibinfo
  {author} {\bibfnamefont {K.}~\bibnamefont {Held}},\ }\href {\doibase
  10.1103/PhysRevB.93.165103} {\bibfield  {journal} {\bibinfo  {journal} {Phys.
  Rev. B}\ }\textbf {\bibinfo {volume} {93}},\ \bibinfo {pages} {165103}
  (\bibinfo {year} {2016})}\BibitemShut {NoStop}%
\bibitem [{\citenamefont {Abrikosov}(1965)}]{Abrikosov1965}%
  \BibitemOpen
  \bibfield  {author} {\bibinfo {author} {\bibfnamefont {A.~A.}\ \bibnamefont
  {Abrikosov}},\ }\href@noop {} {\bibfield  {journal} {\bibinfo  {journal}
  {Physics}\ }\textbf {\bibinfo {volume} {2}},\ \bibinfo {pages} {5} (\bibinfo
  {year} {1965})}\BibitemShut {NoStop}%
\bibitem [{Note1()}]{Note1}%
  \BibitemOpen
  \bibinfo {note} {Our nomenclature follows the seminal application of the
  parquet equations to the X-ray-edge singularity by Roulet et al.\ \cite
  {Roulet1969}. While we use $\Gamma $, $R$, $\gamma _r$, and $I_r$ for the
  full, totally irreducible, two-particle-reducible, and -irreducible vertices,
  respectively, another common choice \cite
  {Rohringer2017,Rohringer2012,Wentzell2016} is given by $F$, $\Lambda $, $\Phi
  _r$, $\Gamma _r$, respectively. Similarly, a common notation \cite
  {Rohringer2017,Rohringer2012,Wentzell2016} for the channels $a, p, t$ is $ph,
  pp, \protect \overline {ph}$, referring to the (longitudinal) particle-hole,
  the particle-particle, and the transverse (or vertical) particle-hole
  channel, respectively. One also finds the labels $x, p, d$ in the literature
  \cite {Jakobs2010}, referring to the so-called exchange, pairing, and direct
  channel, respectively.}\BibitemShut {Stop}%
\bibitem [{\citenamefont {Rohringer}\ \emph {et~al.}(2018)\citenamefont
  {Rohringer}, \citenamefont {Hafermann}, \citenamefont {Toschi}, \citenamefont
  {Katanin}, \citenamefont {Antipov}, \citenamefont {Katsnelson}, \citenamefont
  {Lichtenstein}, \citenamefont {Rubtsov},\ and\ \citenamefont
  {Held}}]{Rohringer2017}%
  \BibitemOpen
  \bibfield  {author} {\bibinfo {author} {\bibfnamefont {G.}~\bibnamefont
  {Rohringer}}, \bibinfo {author} {\bibfnamefont {H.}~\bibnamefont
  {Hafermann}}, \bibinfo {author} {\bibfnamefont {A.}~\bibnamefont {Toschi}},
  \bibinfo {author} {\bibfnamefont {A.~A.}\ \bibnamefont {Katanin}}, \bibinfo
  {author} {\bibfnamefont {A.~E.}\ \bibnamefont {Antipov}}, \bibinfo {author}
  {\bibfnamefont {M.~I.}\ \bibnamefont {Katsnelson}}, \bibinfo {author}
  {\bibfnamefont {A.~I.}\ \bibnamefont {Lichtenstein}}, \bibinfo {author}
  {\bibfnamefont {A.~N.}\ \bibnamefont {Rubtsov}}, \ and\ \bibinfo {author}
  {\bibfnamefont {K.}~\bibnamefont {Held}},\ }\href {\doibase
  10.1103/RevModPhys.90.025003} {\bibfield  {journal} {\bibinfo  {journal}
  {Rev. Mod. Phys.}\ }\textbf {\bibinfo {volume} {90}},\ \bibinfo {pages}
  {025003} (\bibinfo {year} {2018})}\BibitemShut {NoStop}%
\bibitem [{\citenamefont {Rohringer}\ \emph {et~al.}(2012)\citenamefont
  {Rohringer}, \citenamefont {Valli},\ and\ \citenamefont
  {Toschi}}]{Rohringer2012}%
  \BibitemOpen
  \bibfield  {author} {\bibinfo {author} {\bibfnamefont {G.}~\bibnamefont
  {Rohringer}}, \bibinfo {author} {\bibfnamefont {A.}~\bibnamefont {Valli}}, \
  and\ \bibinfo {author} {\bibfnamefont {A.}~\bibnamefont {Toschi}},\ }\href
  {\doibase 10.1103/PhysRevB.86.125114} {\bibfield  {journal} {\bibinfo
  {journal} {Phys. Rev. B}\ }\textbf {\bibinfo {volume} {86}},\ \bibinfo
  {pages} {125114} (\bibinfo {year} {2012})}\BibitemShut {NoStop}%
\bibitem [{\citenamefont {{Wentzell}}\ \emph {et~al.}()\citenamefont
  {{Wentzell}}, \citenamefont {{Li}}, \citenamefont {{Tagliavini}},
  \citenamefont {{Taranto}}, \citenamefont {{Rohringer}}, \citenamefont
  {{Held}}, \citenamefont {{Toschi}},\ and\ \citenamefont
  {{Andergassen}}}]{Wentzell2016}%
  \BibitemOpen
  \bibfield  {author} {\bibinfo {author} {\bibfnamefont {N.}~\bibnamefont
  {{Wentzell}}}, \bibinfo {author} {\bibfnamefont {G.}~\bibnamefont {{Li}}},
  \bibinfo {author} {\bibfnamefont {A.}~\bibnamefont {{Tagliavini}}}, \bibinfo
  {author} {\bibfnamefont {C.}~\bibnamefont {{Taranto}}}, \bibinfo {author}
  {\bibfnamefont {G.}~\bibnamefont {{Rohringer}}}, \bibinfo {author}
  {\bibfnamefont {K.}~\bibnamefont {{Held}}}, \bibinfo {author} {\bibfnamefont
  {A.}~\bibnamefont {{Toschi}}}, \ and\ \bibinfo {author} {\bibfnamefont
  {S.}~\bibnamefont {{Andergassen}}},\ }\href {http://arxiv.org/abs/1610.06520}
  {\ }\Eprint {http://arxiv.org/abs/1610.06520} {arXiv:1610.06520} \BibitemShut
  {NoStop}%
\bibitem [{\citenamefont {Jakobs}\ \emph
  {et~al.}(2010{\natexlab{a}})\citenamefont {Jakobs}, \citenamefont
  {Pletyukhov},\ and\ \citenamefont {Schoeller}}]{Jakobs2010}%
  \BibitemOpen
  \bibfield  {author} {\bibinfo {author} {\bibfnamefont {S.~G.}\ \bibnamefont
  {Jakobs}}, \bibinfo {author} {\bibfnamefont {M.}~\bibnamefont {Pletyukhov}},
  \ and\ \bibinfo {author} {\bibfnamefont {H.}~\bibnamefont {Schoeller}},\
  }\href {\doibase 10.1103/PhysRevB.81.195109} {\bibfield  {journal} {\bibinfo
  {journal} {Phys. Rev. B}\ }\textbf {\bibinfo {volume} {81}},\ \bibinfo
  {pages} {195109} (\bibinfo {year} {2010}{\natexlab{a}})}\BibitemShut
  {NoStop}%
\bibitem [{Note2()}]{Note2}%
  \BibitemOpen
  \bibinfo {note} {Whereas the initial condition $\Gamma ^{\Lambda _i}=R$ at
  $G^{\Lambda _i}=0$ is natural in the parquet approximation $R=\Gamma _0$, it
  might seem counter-intuitive for other cases, when thinking of the totally
  irreducible vertex, $R$, being itself composed of diagrams containing
  propagators. In this way of thinking, we have to treat propagators in $R$
  differently from those propagators that connect the building block $R$ in the
  two-particle-reducible diagrams of the $\gamma _r$. This special treatment is
  necessary as $R$ does not have an efficient flow equation.}\BibitemShut
  {Stop}%
\bibitem [{\citenamefont {Wentzell}\ \emph {et~al.}(2015)\citenamefont
  {Wentzell}, \citenamefont {Taranto}, \citenamefont {Katanin}, \citenamefont
  {Toschi},\ and\ \citenamefont {Andergassen}}]{Wentzell2015}%
  \BibitemOpen
  \bibfield  {author} {\bibinfo {author} {\bibfnamefont {N.}~\bibnamefont
  {Wentzell}}, \bibinfo {author} {\bibfnamefont {C.}~\bibnamefont {Taranto}},
  \bibinfo {author} {\bibfnamefont {A.}~\bibnamefont {Katanin}}, \bibinfo
  {author} {\bibfnamefont {A.}~\bibnamefont {Toschi}}, \ and\ \bibinfo {author}
  {\bibfnamefont {S.}~\bibnamefont {Andergassen}},\ }\href {\doibase
  10.1103/PhysRevB.91.045120} {\bibfield  {journal} {\bibinfo  {journal} {Phys.
  Rev. B}\ }\textbf {\bibinfo {volume} {91}},\ \bibinfo {pages} {045120}
  (\bibinfo {year} {2015})}\BibitemShut {NoStop}%
\bibitem [{\citenamefont {Georges}\ \emph {et~al.}(1996)\citenamefont
  {Georges}, \citenamefont {Kotliar}, \citenamefont {Krauth},\ and\
  \citenamefont {Rozenberg}}]{Georges1996}%
  \BibitemOpen
  \bibfield  {author} {\bibinfo {author} {\bibfnamefont {A.}~\bibnamefont
  {Georges}}, \bibinfo {author} {\bibfnamefont {G.}~\bibnamefont {Kotliar}},
  \bibinfo {author} {\bibfnamefont {W.}~\bibnamefont {Krauth}}, \ and\ \bibinfo
  {author} {\bibfnamefont {M.~J.}\ \bibnamefont {Rozenberg}},\ }\href {\doibase
  10.1103/RevModPhys.68.13} {\bibfield  {journal} {\bibinfo  {journal} {Rev.
  Mod. Phys.}\ }\textbf {\bibinfo {volume} {68}},\ \bibinfo {pages} {13}
  (\bibinfo {year} {1996})}\BibitemShut {NoStop}%
\bibitem [{\citenamefont {Taranto}\ \emph {et~al.}(2014)\citenamefont
  {Taranto}, \citenamefont {Andergassen}, \citenamefont {Bauer}, \citenamefont
  {Held}, \citenamefont {Katanin}, \citenamefont {Metzner}, \citenamefont
  {Rohringer},\ and\ \citenamefont {Toschi}}]{Taranto2014}%
  \BibitemOpen
  \bibfield  {author} {\bibinfo {author} {\bibfnamefont {C.}~\bibnamefont
  {Taranto}}, \bibinfo {author} {\bibfnamefont {S.}~\bibnamefont
  {Andergassen}}, \bibinfo {author} {\bibfnamefont {J.}~\bibnamefont {Bauer}},
  \bibinfo {author} {\bibfnamefont {K.}~\bibnamefont {Held}}, \bibinfo {author}
  {\bibfnamefont {A.}~\bibnamefont {Katanin}}, \bibinfo {author} {\bibfnamefont
  {W.}~\bibnamefont {Metzner}}, \bibinfo {author} {\bibfnamefont
  {G.}~\bibnamefont {Rohringer}}, \ and\ \bibinfo {author} {\bibfnamefont
  {A.}~\bibnamefont {Toschi}},\ }\href {\doibase
  10.1103/PhysRevLett.112.196402} {\bibfield  {journal} {\bibinfo  {journal}
  {Phys. Rev. Lett.}\ }\textbf {\bibinfo {volume} {112}},\ \bibinfo {pages}
  {196402} (\bibinfo {year} {2014})}\BibitemShut {NoStop}%
\bibitem [{\citenamefont {Tagliavini}(2018)}]{Tagliavini2018a}%
  \BibitemOpen
  \bibfield  {author} {\bibinfo {author} {\bibfnamefont {A.}~\bibnamefont
  {Tagliavini}},\ }\href@noop {} {Ph.D. thesis},\ \bibinfo  {school}
  {Universit\"at T\"ubingen} (\bibinfo {year} {2018})\BibitemShut {NoStop}%
\bibitem [{\citenamefont {Toschi}\ \emph {et~al.}(2007)\citenamefont {Toschi},
  \citenamefont {Katanin},\ and\ \citenamefont {Held}}]{Toschi2007}%
  \BibitemOpen
  \bibfield  {author} {\bibinfo {author} {\bibfnamefont {A.}~\bibnamefont
  {Toschi}}, \bibinfo {author} {\bibfnamefont {A.~A.}\ \bibnamefont {Katanin}},
  \ and\ \bibinfo {author} {\bibfnamefont {K.}~\bibnamefont {Held}},\ }\href
  {\doibase 10.1103/PhysRevB.75.045118} {\bibfield  {journal} {\bibinfo
  {journal} {Phys. Rev. B}\ }\textbf {\bibinfo {volume} {75}},\ \bibinfo
  {pages} {045118} (\bibinfo {year} {2007})}\BibitemShut {NoStop}%
\bibitem [{\citenamefont {Held}\ \emph {et~al.}(2008)\citenamefont {Held},
  \citenamefont {Katanin},\ and\ \citenamefont {Toschi}}]{Held2008}%
  \BibitemOpen
  \bibfield  {author} {\bibinfo {author} {\bibfnamefont {K.}~\bibnamefont
  {Held}}, \bibinfo {author} {\bibfnamefont {A.~A.}\ \bibnamefont {Katanin}}, \
  and\ \bibinfo {author} {\bibfnamefont {A.}~\bibnamefont {Toschi}},\ }\href
  {\doibase 10.1143/PTPS.176.117} {\bibfield  {journal} {\bibinfo  {journal}
  {Prog. Theor. Phys. Supp.}\ }\textbf {\bibinfo {volume} {176}},\ \bibinfo
  {pages} {117} (\bibinfo {year} {2008})}\BibitemShut {NoStop}%
\bibitem [{\citenamefont {Valli}\ \emph {et~al.}(2015)\citenamefont {Valli},
  \citenamefont {Sch\"afer}, \citenamefont {Thunstr\"om}, \citenamefont
  {Rohringer}, \citenamefont {Andergassen}, \citenamefont {Sangiovanni},
  \citenamefont {Held},\ and\ \citenamefont {Toschi}}]{Valli2015}%
  \BibitemOpen
  \bibfield  {author} {\bibinfo {author} {\bibfnamefont {A.}~\bibnamefont
  {Valli}}, \bibinfo {author} {\bibfnamefont {T.}~\bibnamefont {Sch\"afer}},
  \bibinfo {author} {\bibfnamefont {P.}~\bibnamefont {Thunstr\"om}}, \bibinfo
  {author} {\bibfnamefont {G.}~\bibnamefont {Rohringer}}, \bibinfo {author}
  {\bibfnamefont {S.}~\bibnamefont {Andergassen}}, \bibinfo {author}
  {\bibfnamefont {G.}~\bibnamefont {Sangiovanni}}, \bibinfo {author}
  {\bibfnamefont {K.}~\bibnamefont {Held}}, \ and\ \bibinfo {author}
  {\bibfnamefont {A.}~\bibnamefont {Toschi}},\ }\href {\doibase
  10.1103/PhysRevB.91.115115} {\bibfield  {journal} {\bibinfo  {journal} {Phys.
  Rev. B}\ }\textbf {\bibinfo {volume} {91}},\ \bibinfo {pages} {115115}
  (\bibinfo {year} {2015})}\BibitemShut {NoStop}%
\bibitem [{\citenamefont {Sch\"afer}\ \emph {et~al.}(2013)\citenamefont
  {Sch\"afer}, \citenamefont {Rohringer}, \citenamefont {Gunnarsson},
  \citenamefont {Ciuchi}, \citenamefont {Sangiovanni},\ and\ \citenamefont
  {Toschi}}]{Schaefer2013}%
  \BibitemOpen
  \bibfield  {author} {\bibinfo {author} {\bibfnamefont {T.}~\bibnamefont
  {Sch\"afer}}, \bibinfo {author} {\bibfnamefont {G.}~\bibnamefont
  {Rohringer}}, \bibinfo {author} {\bibfnamefont {O.}~\bibnamefont
  {Gunnarsson}}, \bibinfo {author} {\bibfnamefont {S.}~\bibnamefont {Ciuchi}},
  \bibinfo {author} {\bibfnamefont {G.}~\bibnamefont {Sangiovanni}}, \ and\
  \bibinfo {author} {\bibfnamefont {A.}~\bibnamefont {Toschi}},\ }\href
  {\doibase 10.1103/PhysRevLett.110.246405} {\bibfield  {journal} {\bibinfo
  {journal} {Phys. Rev. Lett.}\ }\textbf {\bibinfo {volume} {110}},\ \bibinfo
  {pages} {246405} (\bibinfo {year} {2013})}\BibitemShut {NoStop}%
\bibitem [{\citenamefont {Sch\"afer}\ \emph {et~al.}(2016)\citenamefont
  {Sch\"afer}, \citenamefont {Ciuchi}, \citenamefont {Wallerberger},
  \citenamefont {Thunstr\"om}, \citenamefont {Gunnarsson}, \citenamefont
  {Sangiovanni}, \citenamefont {Rohringer},\ and\ \citenamefont
  {Toschi}}]{Schaefer2016}%
  \BibitemOpen
  \bibfield  {author} {\bibinfo {author} {\bibfnamefont {T.}~\bibnamefont
  {Sch\"afer}}, \bibinfo {author} {\bibfnamefont {S.}~\bibnamefont {Ciuchi}},
  \bibinfo {author} {\bibfnamefont {M.}~\bibnamefont {Wallerberger}}, \bibinfo
  {author} {\bibfnamefont {P.}~\bibnamefont {Thunstr\"om}}, \bibinfo {author}
  {\bibfnamefont {O.}~\bibnamefont {Gunnarsson}}, \bibinfo {author}
  {\bibfnamefont {G.}~\bibnamefont {Sangiovanni}}, \bibinfo {author}
  {\bibfnamefont {G.}~\bibnamefont {Rohringer}}, \ and\ \bibinfo {author}
  {\bibfnamefont {A.}~\bibnamefont {Toschi}},\ }\href {\doibase
  10.1103/PhysRevB.94.235108} {\bibfield  {journal} {\bibinfo  {journal} {Phys.
  Rev. B}\ }\textbf {\bibinfo {volume} {94}},\ \bibinfo {pages} {235108}
  (\bibinfo {year} {2016})}\BibitemShut {NoStop}%
\bibitem [{\citenamefont {Gunnarsson}\ \emph {et~al.}(2017)\citenamefont
  {Gunnarsson}, \citenamefont {Rohringer}, \citenamefont {Sch\"afer},
  \citenamefont {Sangiovanni},\ and\ \citenamefont {Toschi}}]{Gunnarsson2017}%
  \BibitemOpen
  \bibfield  {author} {\bibinfo {author} {\bibfnamefont {O.}~\bibnamefont
  {Gunnarsson}}, \bibinfo {author} {\bibfnamefont {G.}~\bibnamefont
  {Rohringer}}, \bibinfo {author} {\bibfnamefont {T.}~\bibnamefont
  {Sch\"afer}}, \bibinfo {author} {\bibfnamefont {G.}~\bibnamefont
  {Sangiovanni}}, \ and\ \bibinfo {author} {\bibfnamefont {A.}~\bibnamefont
  {Toschi}},\ }\href {\doibase 10.1103/PhysRevLett.119.056402} {\bibfield
  {journal} {\bibinfo  {journal} {Phys. Rev. Lett.}\ }\textbf {\bibinfo
  {volume} {119}},\ \bibinfo {pages} {056402} (\bibinfo {year}
  {2017})}\BibitemShut {NoStop}%
\bibitem [{\citenamefont {Chalupa}\ \emph {et~al.}(2018)\citenamefont
  {Chalupa}, \citenamefont {Gunacker}, \citenamefont {Sch\"afer}, \citenamefont
  {Held},\ and\ \citenamefont {Toschi}}]{Chalupa2018}%
  \BibitemOpen
  \bibfield  {author} {\bibinfo {author} {\bibfnamefont {P.}~\bibnamefont
  {Chalupa}}, \bibinfo {author} {\bibfnamefont {P.}~\bibnamefont {Gunacker}},
  \bibinfo {author} {\bibfnamefont {T.}~\bibnamefont {Sch\"afer}}, \bibinfo
  {author} {\bibfnamefont {K.}~\bibnamefont {Held}}, \ and\ \bibinfo {author}
  {\bibfnamefont {A.}~\bibnamefont {Toschi}},\ }\href {\doibase
  10.1103/PhysRevB.97.245136} {\bibfield  {journal} {\bibinfo  {journal} {Phys.
  Rev. B}\ }\textbf {\bibinfo {volume} {97}},\ \bibinfo {pages} {245136}
  (\bibinfo {year} {2018})}\BibitemShut {NoStop}%
\bibitem [{\citenamefont {Thunstr\"om}\ \emph {et~al.}(2018)\citenamefont
  {Thunstr\"om}, \citenamefont {Gunnarsson}, \citenamefont {Ciuchi},\ and\
  \citenamefont {Rohringer}}]{Thunstroem2018}%
  \BibitemOpen
  \bibfield  {author} {\bibinfo {author} {\bibfnamefont {P.}~\bibnamefont
  {Thunstr\"om}}, \bibinfo {author} {\bibfnamefont {O.}~\bibnamefont
  {Gunnarsson}}, \bibinfo {author} {\bibfnamefont {S.}~\bibnamefont {Ciuchi}},
  \ and\ \bibinfo {author} {\bibfnamefont {G.}~\bibnamefont {Rohringer}},\
  }\href {\doibase 10.1103/PhysRevB.98.235107} {\bibfield  {journal} {\bibinfo
  {journal} {Phys. Rev. B}\ }\textbf {\bibinfo {volume} {98}},\ \bibinfo
  {pages} {235107} (\bibinfo {year} {2018})}\BibitemShut {NoStop}%
\bibitem [{\citenamefont {Luttinger}\ and\ \citenamefont
  {Ward}(1960)}]{Luttinger1960}%
  \BibitemOpen
  \bibfield  {author} {\bibinfo {author} {\bibfnamefont {J.~M.}\ \bibnamefont
  {Luttinger}}\ and\ \bibinfo {author} {\bibfnamefont {J.~C.}\ \bibnamefont
  {Ward}},\ }\href {\doibase 10.1103/PhysRev.118.1417} {\bibfield  {journal}
  {\bibinfo  {journal} {Phys. Rev.}\ }\textbf {\bibinfo {volume} {118}},\
  \bibinfo {pages} {1417} (\bibinfo {year} {1960})}\BibitemShut {NoStop}%
\bibitem [{\citenamefont {Baym}(1962)}]{Baym1962}%
  \BibitemOpen
  \bibfield  {author} {\bibinfo {author} {\bibfnamefont {G.}~\bibnamefont
  {Baym}},\ }\href {\doibase 10.1103/PhysRev.127.1391} {\bibfield  {journal}
  {\bibinfo  {journal} {Phys. Rev.}\ }\textbf {\bibinfo {volume} {127}},\
  \bibinfo {pages} {1391} (\bibinfo {year} {1962})}\BibitemShut {NoStop}%
\bibitem [{\citenamefont {Katanin}(2004)}]{Katanin2004}%
  \BibitemOpen
  \bibfield  {author} {\bibinfo {author} {\bibfnamefont {A.~A.}\ \bibnamefont
  {Katanin}},\ }\href {\doibase 10.1103/PhysRevB.70.115109} {\bibfield
  {journal} {\bibinfo  {journal} {Phys. Rev. B}\ }\textbf {\bibinfo {volume}
  {70}},\ \bibinfo {pages} {115109} (\bibinfo {year} {2004})}\BibitemShut
  {NoStop}%
\bibitem [{Note3()}]{Note3}%
  \BibitemOpen
  \bibinfo {note} {The substitution $S \to \protect \mathaccentV {dot}05F{G}$
  in the truncated fRG vertex flow is often called Katanin substitution \cite
  {Katanin2004}.}\BibitemShut {Stop}%
\bibitem [{\citenamefont {Kugler}(2018)}]{Kugler2018c}%
  \BibitemOpen
  \bibfield  {author} {\bibinfo {author} {\bibfnamefont {F.~B.}\ \bibnamefont
  {Kugler}},\ }\href {\doibase 10.1103/PhysRevE.98.023303} {\bibfield
  {journal} {\bibinfo  {journal} {Phys. Rev. E}\ }\textbf {\bibinfo {volume}
  {98}},\ \bibinfo {pages} {023303} (\bibinfo {year} {2018})}\BibitemShut
  {NoStop}%
\bibitem [{\citenamefont {Jani\ifmmode~\check{s}\else \v{s}\fi{}}\ \emph
  {et~al.}(2017)\citenamefont {Jani\ifmmode~\check{s}\else \v{s}\fi{}},
  \citenamefont {Kauch},\ and\ \citenamefont {Pokorn\'y}}]{Janis2017}%
  \BibitemOpen
  \bibfield  {author} {\bibinfo {author} {\bibfnamefont {V.}~\bibnamefont
  {Jani\ifmmode~\check{s}\else \v{s}\fi{}}}, \bibinfo {author} {\bibfnamefont
  {A.}~\bibnamefont {Kauch}}, \ and\ \bibinfo {author} {\bibfnamefont
  {V.}~\bibnamefont {Pokorn\'y}},\ }\href {\doibase 10.1103/PhysRevB.95.045108}
  {\bibfield  {journal} {\bibinfo  {journal} {Phys. Rev. B}\ }\textbf {\bibinfo
  {volume} {95}},\ \bibinfo {pages} {045108} (\bibinfo {year}
  {2017})}\BibitemShut {NoStop}%
\bibitem [{\citenamefont {Kugler}\ and\ \citenamefont {von
  Delft}(2018{\natexlab{c}})}]{Kugler2018b}%
  \BibitemOpen
  \bibfield  {author} {\bibinfo {author} {\bibfnamefont {F.~B.}\ \bibnamefont
  {Kugler}}\ and\ \bibinfo {author} {\bibfnamefont {J.}~\bibnamefont {von
  Delft}},\ }\href {http://stacks.iop.org/0953-8984/30/i=19/a=195501}
  {\bibfield  {journal} {\bibinfo  {journal} {J. Phys.: Condens. Matter}\
  }\textbf {\bibinfo {volume} {30}},\ \bibinfo {pages} {195501} (\bibinfo
  {year} {2018}{\natexlab{c}})}\BibitemShut {NoStop}%
\bibitem [{\citenamefont {Eberlein}(2014)}]{Eberlein2014}%
  \BibitemOpen
  \bibfield  {author} {\bibinfo {author} {\bibfnamefont {A.}~\bibnamefont
  {Eberlein}},\ }\href {\doibase 10.1103/PhysRevB.90.115125} {\bibfield
  {journal} {\bibinfo  {journal} {Phys. Rev. B}\ }\textbf {\bibinfo {volume}
  {90}},\ \bibinfo {pages} {115125} (\bibinfo {year} {2014})}\BibitemShut
  {NoStop}%
\bibitem [{\citenamefont {R\"uck}\ and\ \citenamefont
  {Reuther}(2018)}]{Rueck2018}%
  \BibitemOpen
  \bibfield  {author} {\bibinfo {author} {\bibfnamefont {M.}~\bibnamefont
  {R\"uck}}\ and\ \bibinfo {author} {\bibfnamefont {J.}~\bibnamefont
  {Reuther}},\ }\href {\doibase 10.1103/PhysRevB.97.144404} {\bibfield
  {journal} {\bibinfo  {journal} {Phys. Rev. B}\ }\textbf {\bibinfo {volume}
  {97}},\ \bibinfo {pages} {144404} (\bibinfo {year} {2018})}\BibitemShut
  {NoStop}%
\bibitem [{\citenamefont {Mitra}\ \emph {et~al.}(1987)\citenamefont {Mitra},
  \citenamefont {Chatterjee},\ and\ \citenamefont {Mukhopadhyay}}]{Mitra1987}%
  \BibitemOpen
  \bibfield  {author} {\bibinfo {author} {\bibfnamefont {T.}~\bibnamefont
  {Mitra}}, \bibinfo {author} {\bibfnamefont {A.}~\bibnamefont {Chatterjee}}, \
  and\ \bibinfo {author} {\bibfnamefont {S.}~\bibnamefont {Mukhopadhyay}},\
  }\href {\doibase https://doi.org/10.1016/0370-1573(87)90087-1} {\bibfield
  {journal} {\bibinfo  {journal} {Phys. Rep.}\ }\textbf {\bibinfo {volume}
  {153}},\ \bibinfo {pages} {91 } (\bibinfo {year} {1987})}\BibitemShut
  {NoStop}%
\bibitem [{\citenamefont {Schmidt}\ \emph {et~al.}(2018)\citenamefont
  {Schmidt}, \citenamefont {Knap}, \citenamefont {Ivanov}, \citenamefont {You},
  \citenamefont {Cetina},\ and\ \citenamefont {Demler}}]{Schmidt2018}%
  \BibitemOpen
  \bibfield  {author} {\bibinfo {author} {\bibfnamefont {R.}~\bibnamefont
  {Schmidt}}, \bibinfo {author} {\bibfnamefont {M.}~\bibnamefont {Knap}},
  \bibinfo {author} {\bibfnamefont {D.~A.}\ \bibnamefont {Ivanov}}, \bibinfo
  {author} {\bibfnamefont {J.-S.}\ \bibnamefont {You}}, \bibinfo {author}
  {\bibfnamefont {M.}~\bibnamefont {Cetina}}, \ and\ \bibinfo {author}
  {\bibfnamefont {E.}~\bibnamefont {Demler}},\ }\href
  {http://stacks.iop.org/0034-4885/81/i=2/a=024401} {\bibfield  {journal}
  {\bibinfo  {journal} {Rep. Prog. Phys.}\ }\textbf {\bibinfo {volume} {81}},\
  \bibinfo {pages} {024401} (\bibinfo {year} {2018})}\BibitemShut {NoStop}%
\bibitem [{\citenamefont {Reuther}\ and\ \citenamefont
  {W\"olfle}(2010)}]{Reuther2010}%
  \BibitemOpen
  \bibfield  {author} {\bibinfo {author} {\bibfnamefont {J.}~\bibnamefont
  {Reuther}}\ and\ \bibinfo {author} {\bibfnamefont {P.}~\bibnamefont
  {W\"olfle}},\ }\href {\doibase 10.1103/PhysRevB.81.144410} {\bibfield
  {journal} {\bibinfo  {journal} {Phys. Rev. B}\ }\textbf {\bibinfo {volume}
  {81}},\ \bibinfo {pages} {144410} (\bibinfo {year} {2010})}\BibitemShut
  {NoStop}%
\bibitem [{\citenamefont {Schmidt}\ and\ \citenamefont
  {Enss}(2011)}]{Schmidt2011}%
  \BibitemOpen
  \bibfield  {author} {\bibinfo {author} {\bibfnamefont {R.}~\bibnamefont
  {Schmidt}}\ and\ \bibinfo {author} {\bibfnamefont {T.}~\bibnamefont {Enss}},\
  }\href {\doibase 10.1103/PhysRevA.83.063620} {\bibfield  {journal} {\bibinfo
  {journal} {Phys. Rev. A}\ }\textbf {\bibinfo {volume} {83}},\ \bibinfo
  {pages} {063620} (\bibinfo {year} {2011})}\BibitemShut {NoStop}%
\bibitem [{\citenamefont {Meden}\ \emph {et~al.}(2008)\citenamefont {Meden},
  \citenamefont {Andergassen}, \citenamefont {Enss}, \citenamefont
  {Schoeller},\ and\ \citenamefont {Sch\"onhammer}}]{Meden2008}%
  \BibitemOpen
  \bibfield  {author} {\bibinfo {author} {\bibfnamefont {V.}~\bibnamefont
  {Meden}}, \bibinfo {author} {\bibfnamefont {S.}~\bibnamefont {Andergassen}},
  \bibinfo {author} {\bibfnamefont {T.}~\bibnamefont {Enss}}, \bibinfo {author}
  {\bibfnamefont {H.}~\bibnamefont {Schoeller}}, \ and\ \bibinfo {author}
  {\bibfnamefont {K.}~\bibnamefont {Sch\"onhammer}},\ }\href
  {http://stacks.iop.org/1367-2630/10/i=4/a=045012} {\bibfield  {journal}
  {\bibinfo  {journal} {New J.\ Phys.}\ }\textbf {\bibinfo {volume} {10}},\
  \bibinfo {pages} {045012} (\bibinfo {year} {2008})}\BibitemShut {NoStop}%
\bibitem [{\citenamefont {Imambekov}\ \emph {et~al.}(2012)\citenamefont
  {Imambekov}, \citenamefont {Schmidt},\ and\ \citenamefont
  {Glazman}}]{Imambekov2012}%
  \BibitemOpen
  \bibfield  {author} {\bibinfo {author} {\bibfnamefont {A.}~\bibnamefont
  {Imambekov}}, \bibinfo {author} {\bibfnamefont {T.~L.}\ \bibnamefont
  {Schmidt}}, \ and\ \bibinfo {author} {\bibfnamefont {L.~I.}\ \bibnamefont
  {Glazman}},\ }\href {\doibase 10.1103/RevModPhys.84.1253} {\bibfield
  {journal} {\bibinfo  {journal} {Rev. Mod. Phys.}\ }\textbf {\bibinfo {volume}
  {84}},\ \bibinfo {pages} {1253} (\bibinfo {year} {2012})}\BibitemShut
  {NoStop}%
\bibitem [{\citenamefont {Abanin}\ and\ \citenamefont
  {Papi\'c}(2017)}]{Abanin2017}%
  \BibitemOpen
  \bibfield  {author} {\bibinfo {author} {\bibfnamefont {D.~A.}\ \bibnamefont
  {Abanin}}\ and\ \bibinfo {author} {\bibfnamefont {Z.}~\bibnamefont
  {Papi\'c}},\ }\href {\doibase 10.1002/andp.201700169} {\bibfield  {journal}
  {\bibinfo  {journal} {Ann. Phys.}\ }\textbf {\bibinfo {volume} {529}},\
  \bibinfo {pages} {1700169} (\bibinfo {year} {2017})}\BibitemShut {NoStop}%
\bibitem [{\citenamefont {Pietracaprina}\ \emph {et~al.}(2018)\citenamefont
  {Pietracaprina}, \citenamefont {Macé}, \citenamefont {Luitz},\ and\
  \citenamefont {Alet}}]{Pietracaprina2018}%
  \BibitemOpen
  \bibfield  {author} {\bibinfo {author} {\bibfnamefont {F.}~\bibnamefont
  {Pietracaprina}}, \bibinfo {author} {\bibfnamefont {N.}~\bibnamefont
  {Macé}}, \bibinfo {author} {\bibfnamefont {D.~J.}\ \bibnamefont {Luitz}}, \
  and\ \bibinfo {author} {\bibfnamefont {F.}~\bibnamefont {Alet}},\ }\href
  {\doibase 10.21468/SciPostPhys.5.5.045} {\bibfield  {journal} {\bibinfo
  {journal} {SciPost Phys.}\ }\textbf {\bibinfo {volume} {5}},\ \bibinfo
  {pages} {45} (\bibinfo {year} {2018})}\BibitemShut {NoStop}%
\bibitem [{\citenamefont {Amini}\ \emph {et~al.}(2014)\citenamefont {Amini},
  \citenamefont {Kravtsov},\ and\ \citenamefont {M\"uller}}]{Amini2014}%
  \BibitemOpen
  \bibfield  {author} {\bibinfo {author} {\bibfnamefont {M.}~\bibnamefont
  {Amini}}, \bibinfo {author} {\bibfnamefont {V.~E.}\ \bibnamefont {Kravtsov}},
  \ and\ \bibinfo {author} {\bibfnamefont {M.}~\bibnamefont {M\"uller}},\
  }\href {http://stacks.iop.org/1367-2630/16/i=1/a=015022} {\bibfield
  {journal} {\bibinfo  {journal} {New J. Phys.}\ }\textbf {\bibinfo {volume}
  {16}},\ \bibinfo {pages} {015022} (\bibinfo {year} {2014})}\BibitemShut
  {NoStop}%
\bibitem [{\citenamefont {Jakobs}\ \emph
  {et~al.}(2010{\natexlab{b}})\citenamefont {Jakobs}, \citenamefont
  {Pletyukhov},\ and\ \citenamefont {Schoeller}}]{Jakobs2010a}%
  \BibitemOpen
  \bibfield  {author} {\bibinfo {author} {\bibfnamefont {S.~G.}\ \bibnamefont
  {Jakobs}}, \bibinfo {author} {\bibfnamefont {M.}~\bibnamefont {Pletyukhov}},
  \ and\ \bibinfo {author} {\bibfnamefont {H.}~\bibnamefont {Schoeller}},\
  }\href {http://stacks.iop.org/1751-8121/43/i=10/a=103001} {\bibfield
  {journal} {\bibinfo  {journal} {J. Phys. A}\ }\textbf {\bibinfo {volume}
  {43}},\ \bibinfo {pages} {103001} (\bibinfo {year}
  {2010}{\natexlab{b}})}\BibitemShut {NoStop}%
\end{thebibliography}%

\end{document}